\documentclass[iop,apj,twocolappendix]{emulateapj}

\pdfoutput=1

\usepackage{apjfonts}
\usepackage[UKenglish]{isodate}

\cleanlookdateon

\defcitealias{bellini2014}{Paper 1}
\defcitealias{watkins2015}{Paper 2}
\defcitealias{harris1996}{H96}

\newcommand {\wcen} {$\omega$~Centauri}
\newcommand {\kms} {km\,s$^{-1}$}
\newcommand {\masyr} {mas\,yr$^{-1}$}
\newcommand {\muasyr} {$\mu$as\,yr$^{-1}$}
\newcommand {\Msun} {M$_\odot$}
\newcommand {\Lsun} {L$_\odot$}

\def \Nused {N_\mathrm{used}}
\def \Nfound {N_\mathrm{found}}
\def \Nuf {\Nused / \Nfound}

\usepackage[breaklinks,colorlinks,citecolor=blue,linkcolor=blue,urlcolor=blue,pdftex,bookmarks=true]{hyperref}
\usepackage[all]{hypcap}

\hypersetup{
    pdftitle={HSTPROMO GCs. III. Dynamical distances and mass-to-light ratios},
    pdfauthor={Laura L. Watkins, et al.}
}

\def\equationautorefname~#1\null{%
  equation~(#1)\null
}

\begin{document}

\title{\textit{Hubble Space Telescope} proper motion (HSTPROMO) catalogs of Galactic globular clusters$^{\ast}$. III. Dynamical distances and mass-to-light ratios}

\altaffiltext{$\ast$}{Based on proprietary and archival observations with the NASA/ESA \textit{Hubble Space Telescope}, obtained at the Space Telescope Science Institute, which is operated by AURA, Inc., under NASA contract NAS 5-26555.}

\author{Laura~L.~Watkins, Roeland~P.~van~der~Marel, Andrea~Bellini, Jay~Anderson}
\affil{Space Telescope Science Institute, 3700 San Martin Drive, Baltimore MD 21218, USA}
\email{lwatkins@stsci.edu}

\slugcomment{Draft version, \today, Accepted for publication}
\shorttitle{HSTPROMO GCs. III. Dynamical distances and mass-to-light ratios}
\shortauthors{L.~L.~Watkins~et~al.}


\begin{abstract}
    
    We present dynamical distance estimates for 15 Galactic globular clusters and use these to check the consistency of dynamical and photometric distance estimates. For most of the clusters, this is the first dynamical distance estimate ever determined. We extract proper-motion dispersion profiles using cleaned samples of bright stars from the \textit{Hubble Space Telescope} proper-motion catalogs recently presented in \citet{bellini2014} and compile a set of line-of-sight velocity-dispersion profiles from a variety of literature sources. Distances are then estimated by fitting spherical, non-rotating, isotropic, constant mass-to-light (M/L) dynamical models to the proper-motion and line-of-sight dispersion profiles together. We compare our dynamical distance estimates with literature photometric estimates from the \citet[][2010 edition]{harris1996} globular cluster catalog and find that the mean fractional difference between the two types is consistent with zero at just $-1.9 \pm 1.7 \%$. This indicates that there are no significant biases in either estimation method and provides an important validation of the stellar-evolution theory that underlies photometric distance estimates. The analysis also estimates dynamical M/L ratios for our clusters; on average, the dynamically-inferred M/L ratios agree with existing stellar-population-based M/L ratios that assume a Chabrier initial mass function (IMF) to within $-8.8 \pm 6.4 \%$, implying that such an IMF is consistent with our data. Our results are also consistent with a Kroupa IMF, but strongly rule out a Salpeter IMF. We detect no correlation between our M/L offsets from literature values and our distance offsets from literature values, strongly indicating that our methods are reliable and our results are robust.
   
\end{abstract}



\section{Introduction}
\label{sect:introduction}

Globular clusters (GCs) are among the first objects to have formed in the Universe, within 1~Gyr of the Big Bang. They are some of the oldest objects for which ages are known and, thus, provide crucial constraints on the age of the Universe itself \citep[eg.][]{vandenberg1996, carretta2000, krauss2003}. As \citet{verde2013} pointed out, local measurements of cosmological parameters (such as the age of the Universe) are valuable as they provide cosmology-independent tests of favoured cosmological models.

Three primary methods are used to estimate GC ages: luminosity of the main-sequence turn off \citep[MSTO, eg.][]{chaboyer1995}; white-dwarf (WD) cooling curves \citep{hansen2002}; and nucleocosmochronology \citep[radioactive age dating of stars, eg.][]{cowan2002}. The MSTO luminosity method is the most common, however it requires an accurate determination of the observed magnitude of the MSTO, accurate reddenings, and accurate distances. Advances in photometric instrumentation have provided both the depth and precision necessary to pinpoint the MSTO within cluster CMDs. Reddening maps, such as those presented in \citet{schlafly2011}, fulfil the second requirement. However, cluster distances are still poorly determined and distance uncertainties dominate over other sources of error in age determinations. There are two key points we need to address: 1) we must ensure that we are measuring distances accurately; 2) we must ensure that different distance measures agree (ie. our methods are not biased).

Cluster distances are also useful to obtain for their own merit. They constitute an important rung on the short end of the distance ladder, which can again have implications for verification of cosmological models. Furthermore, distances provide a third dimension of positional phase space. Many clusters are coincident with substructures in the Galactic halo \citep[eg.][]{bellazzini2003, martin2004, mackey2004}; distances to the clusters (and to the substructures) can help us determine if there is simply a chance alignment of cluster and substructure along the LOS or if they are truly associated. Evidence of tidal tails has also been found around some clusters \citep[eg.][]{niedersteostholt2010, kuepper2015} and other clusters have been posited as progenitors for tidal streams of undetermined origin; distance information can help in both cases by providing extra information with which to constrain cluster orbits. Cluster orbits, in turn, can help to constrain the shape of the inner Galactic halo.

Cluster distances can be estimated from both photometric and kinematic data.  Photometric studies typically estimate distances via the distance modulus, defined as the difference between the apparent and absolute magnitudes of a given object or population. Apparent magnitudes are relatively easy to obtain. Absolute magnitudes are only well defined for certain types of objects, often called ``standard candles''. Periodic variables, such as Cepheids and RR Lyraes, are common standard candles in the local Universe, although even their absolute magnitudes do depend on the metallicity of the star and the period of their variability \citep[eg.][]{catelan2004}. Nevertheless, they are very useful distance indicators and are widely used; not only for cluster-distance estimation \citep[eg.][]{delprincipe2006}, but also for many other science applications, including to identify substructure in and trace the shape of the Galactic halo \citep[eg.][]{watkins2009}.

Kinematic studies estimate distances by comparing line-of-sight (LOS) velocity dispersions and proper-motion (PM) velocity dispersions. These quantities are measured in different units; converting from one set of units to the other requires knowledge of the distance. Or, conversely, by assuming that the dispersions are equal (or different in some known and quantifiable way), the distance can be estimated. Very few kinematic distance estimates exist, as only very few clusters have measured internal PMs.

Although, kinematic and photometric distance estimates use very different data sets and very different methods, their estimates should, of course, be the same for any given cluster. Until now, it has been difficult to assess the agreement (or lack thereof) between kinematic and photometric distance estimates due to the scarcity of kinematic distance estimates. However, in the few cases where both photometric and kinematic estimates do exist, they often disagree \citep[eg.][]{delprincipe2006, watkins2013}.

Pinpointing the source of the disagreement is complicated by the fact that both types of estimate are usually made on a cluster-by-cluster basis. So when comparing a set of photometric estimates with a set of kinematic estimates, it is difficult to determine whether any differences are due to systematic biases in either estimation method, or are due to scatter from study to study.

To make progress, we need large, homogeneous sets of distance estimates: larger samples will provide more robust statistics; homogeneity in the distance estimation process will eliminate some potential sources of disagreement so we can be confident we are making a meaningful comparison.

Most GC distances have been measured in individual studies, however, fortunately, there is a large photometric distance analysis that suits our purposes: \citet[][2010 edition, hereafter \citetalias{harris1996}]{harris1996} provided a catalogue of cluster properties (including distances) for 157 Galactic GCs. No large study has yet been done for dynamical distance estimates, as the kinematic data has simply not been available. We require both LOS and PM data; although both types of data are lacking, the dearth of PM data has, so far, been the limiting factor for kinematic distance estimates.

This situation has recently changed. In \citet[][hereafter \citetalias{bellini2014}]{bellini2014}, we presented a set of \textit{Hubble Space Telescope} (\textit{HST}) PM catalogues for 22 GCs in the Milky Way. The catalogues were compiled following a search through archival \textit{HST} data to find fields in Galactic GCs that had been previously observed at multiple epochs for different projects. The exquisite astrometric sensitivity of \textit{HST}, coupled with baselines of up to 12 years (achievable thanks to the archival nature of the project), allowed us to measure PMs with exceptional precision. For typical clusters in our sample, the bright, well-measured stars have measurement uncertainties of just 32~\muasyr\ ($\sim$1.2~\kms\ at a distance of 8~kpc, which is typical for our clusters). Another advantage of these catalogues is the sheer volume of data they encompass: the median number of stars in the full catalogues is $\sim$60,000 and, following certain selection cuts as we will describe in \autoref{sect:data}, we use a median of 3000 stars per cluster for the analysis we present here.

These catalogues were compiled with two key drivers: 1) to facilitate more-detailed studies of each individual cluster; and 2) for the insight we can gain by considering the population of clusters as a whole. There is a wide range of science questions that can be addressed both for individual clusters and for the whole population, too many to consider in detail here, but see the introduction of \citetalias{bellini2014}. Here, we seek to advance both goals.

In \citet[][hereafter \citetalias{watkins2015}]{watkins2015}, we presented kinematical profiles and maps for the bright stars in each of the 22 clusters. Here we build on this analysis by comparing our kinematics with previous LOS studies in order to estimate distances for the clusters. This is the first dynamical distance study performed for more than a handful of clusters, and greatly increases the number of dynamical distances available for Galactic GCs. By comparing these dynamical distances estimates with photometric estimates from the literature, we can assess the consistency of photometric and kinematic distance-estimation methods.

\autoref{sect:data} introduces the PM catalogues, LOS literature data and surface brightness profiles we use for our study. In \autoref{sect:distest}, we briefly outline our distance estimation method and, in \autoref{sect:results}, we use it to estimate distances for our clusters. In \autoref{sect:discussion}, we provide context for our results via comparison with previous studies, and look at statistics for the sample as a whole. We conclude in \autoref{sect:conclusions}.


\section{Cluster data}
\label{sect:data}

To estimate cluster distances, we require both PM and LOS velocity-dispersion profiles. For our analysis, we use the \textit{HST} PM catalogues described in \citetalias{bellini2014}, to which we refer for detailed explanations of the data reduction and processing. We augment these catalogues with LOS velocity data available in the literature. Although \citetalias{bellini2014} provides catalogues for 22 Galactic GCs, here we are only able to use the clusters for which LOS dispersion profiles have been published or for which LOS velocity catalogues have been published containing sufficient stars from which to determine a dispersion profile. Of the 17 clusters with both PM and LOS data, we exclude NGC\,6362 and NGC\,7099 (M\,30) as our PM data sets are too small (following the cuts we make below) for our distance estimation procedure.

As we will see in \autoref{sect:distest}, we also require a Multi-Gaussian-Expansion (MGE)\footnote{MGEs are examples of Finite Mixture Models \citep{mclachlan2000}.} fit to the surface brightness profile for each cluster. Fortunately surface brightness profiles are available for all of our clusters so we do not need to further cut our sample. This leaves us with 15 clusters in total for this study.

Here we briefly describe the PM catalogues, the literature LOS data, the surface brightness profiles we use and the MGE fitting process.

\subsection{Proper-motion catalogues}
\label{sect:pmdata}

GCs are collisional systems; as the stars undergo a series of two-body interactions, they exchange energy and slowly move towards a state of energy equipartition. As a result, low-mass (faint) stars tend to move faster than high-mass (bright) stars. Recently, \citet{trenti2013} showed that clusters never reach full equipartition, but nevertheless, even in partial equipartition, we would still expect that stars of different mass will exhibit different kinematic properties. Thus, to estimate distances via comparison of PM and LOS dispersion profiles, we must ensure that both measurements are for stars of similar mass. LOS studies are typically limited only to the brightest stars; most data is collected from ground-based facilities and only spectra of bright stars achieve sufficient signal-to-noise to measure a velocity. As such, we must impose a magnitude cut on our PM sample to select stars of similar mass to those observed in LOS studies.

Furthermore, unbiased kinematics can only be extracted using stars for which positions (and, hence, PMs) have been accurately measured and for which uncertainties have been correctly estimated. Poor position measurements -- particularly a problem in crowded fields where stars suffer greatly from blending -- and underestimated error bars will both tend to artificially inflate the velocity distributions. The catalogues also contain some contaminants, both from the Milky Way and from other nearby objects, which must be removed.

In \citetalias{watkins2015}, we described a series of cuts that we make to the PM catalogues to ensure that we have a reliable, high-quality sample of stars. We refer to \citetalias{watkins2015} for a detailed discussion of the cuts, but briefly outline the main points here:
\begin{enumerate}
    \item Remove all stars fainter than 1 magnitude below the main-sequence turn off.
    \item Remove all stars with $\Nuf < 0.8$, where $\Nused$ is the number of data points used for the final calculation of the PM and $\Nfound$ is the total number of data points found for each star.
    \item Remove all stars with $F_\mathrm{D} \left( D \chi^2_\mathrm{reduced} \right) > 0.99$, where the $\chi^2_\mathrm{reduced}$ values are for the straight-line fits used to determine PMs, $F_\mathrm{D}$ is the cumulative distribution function for a $\chi^2$ distribution with $D$ degrees of freedom and, here, $D = \Nused - 2$.
    \item Remove stars for which the quality of the point-spread function (PSF) fit was determined to be low. Quality of the PSF fit varies with both magnitude and position within the cluster, and also depends on the distance and concentration of the cluster. As such, the details of this step varied from cluster to cluster.
    \item 3-$\sigma$ clip in velocity and remove all stars with PM errors larger than half the local dispersion. This is an iterative process to ensure that only ``good" stars (ie. measurements with enough signal) are used to define the boundaries of the final PM and PM error cuts.
\end{enumerate}

For \citetalias{watkins2015}, we required a limited range of stellar masses across our sample so that we could neglect the effect of stellar mass on the kinematics. It is this that motivated the initial magnitude cut. A cut 1 magnitude below the main-sequence turn off was adequate for our purpose of calculating spatial kinematic profiles. However, if we are to calculate accurate distances via comparison to LOS data, we must be more restrictive and impose a brighter magnitude cut. So, we take the final samples from \citetalias{watkins2015}, extracted via the cuts listed above, and make a further magnitude cut, retaining only stars brighter than the main-sequence turn off.\footnote{In principle, we can make this revised magnitude cut at the main-sequence turn off in step 1, however this significantly reduces the number of stars in the sample. We then find it difficult to perform the last two stages of the cleaning process effectively due to the nature of the cleaning algorithm. The two-stage magnitude cut we describe in the text is more effective as it limits the mass-range of the initial sample while retaining enough stars with which to perform the rest of the cleaning.}

\subsection{Proper-motion kinematic profiles}
\label{sect:kinematics}

We determine kinematic profiles from our PM data by binning the cleaned samples in radius and then calculating the mean and dispersion of the velocity in each bin. Our PM profiles measure the average 1-dimensional PM in the plane of the sky, which corresponds to an average over the radial and tangential PM components \citep[this is rigorously defined in Section 2.6 of][]{vandermarel2010}.

When binning stars, we need to ensure that we have sufficient stars with which to calculate reliable dispersions, but we also want to maximise the spatial resolution of our bins. As the spatial coverage of the data is highly inhomogenous, we find that a variable binning scheme is best suited to our needs. Then the kinematics in each bin are determined using a maximum-likelihood estimation method that fully accounts for biases. Both the binning scheme and the maximum-likelihood kinematic estimation are described in full in \citetalias{watkins2015}, to which we refer the reader for further details.

As described above, the final PM samples we use for this paper are different from those we used in \citetalias{watkins2015} as we performed an additional magnitude cut for our present study. As a result, the kinematic profiles are not the same as those used in \citetalias{watkins2015}.

\subsection{Line-of-sight velocity and proper-motion literature data}
\label{sect:litdata}

Where possible, we use LOS velocity-dispersion profiles taken directly from previous studies. However, dispersion profiles are not always available; in that case, we use the published stellar catalogue of LOS velocities and determine our own velocity-dispersion profile. To do this, we proceeded similarly as for our PM data: we binned the stars in radius (albeit with a simpler binning scheme\footnote{For the LOS data sets, we separated the stars into $N$ equally-populated bins, where $N$ was chosen separately for each data set; we did not use the variable-binning algorithm used for our PM dispersion profiles.}) and used the same maximum-likelihood method to determine the kinematics in each bin.

\begin{table}
    \caption{Sources for literature data.}
    \label{table:litsource}
    \centering
    
    \begin{tabular}{ccc}
        \hline
        \hline
        Cluster & Source & Type \\
        (1) & (2) & (3) \\
        \hline
        
        NGC\,104 & \citet{gebhardt1995} & LC \\
        & \citet{mclaughlin2006} & LD, PD \\
        & \citet{lane2010} & LD \\
        
        NGC\,288 & \citet{scarpa2007a} & LD \\
        & \citet{sollima2012} & LD \\
        
        NGC\,1851 & \citet{scarpa2011} & LD \\
        & \citet{carretta2011} & LC \\
        & \citet{luetzgendorf2013} & LD \\
        & \citet{lardo2015} & LD \\
        & \citet{larson} & LD \\
        
        NGC\,2808 & \citet{carretta2006} & LC \\
        & \citet{luetzgendorf2012} & LD \\
        & \citet{lardo2015} & LD \\
        & \citet{larson} & LD \\
        
        NGC\,5139 & \citet{vandeven2006} & LD, PD \\
        & \citet{sollima2009} & LD \\
        & \citet{vandermarel2010} & PD \\
        & \citet{noyola2010} & LD \\
        & \citet{scarpa2010} & LD \\
        
        NGC\,5904 & \citet{kimmig2015} & LC \\
        
        NGC\,5927 & \citet{lardo2015} & LD \\
        & \citet{simmerer2013} & LC \\
        
        NGC\,6266 & \citet{mcnamara2012} & PD \\
        & \citet{luetzgendorf2013} & LD \\
        
        NGC\,6341 & \citet{drukier2007} & LC \\
        & \citet{meszaros2009} & LC \\
        & \citet{kamann2014} & LD \\
        & \citet{kimmig2015} & LC \\
        
        NGC\,6388 & \citet{luetzgendorf2011} & LD \\
        & \citet{lanzoni2013} & LD \\
        & \citet{lapenna2015} & LD \\
        
        NGC\,6397 & \citet{meylan1991} & LD \\
        & \citet{gebhardt1995} & LC \\
        & \citet{lovisi2012} & LC \\
        & \citet{heyl2012} & PD \\
        
        NGC\,6656 & \citet{lane2009} & LD \\
        & \citet{zloczewski2012} & LC \\
        
        NGC\,6715 & \citet{bellazzini2008} & LD \\
        & \citet{ibata2009} & LD \\
        & \citet{carretta2010} & LC \\
        
        NGC\,6752 & \citet{carretta2007} & LC \\
        & \citet{sollima2012} & LD \\
        & \citet{zloczewski2012} & LC \\
        & \citet{lardo2015} & LD \\
        
        NGC\,7078 & \citet{drukier1998} & LD \\
        & \citet{mcnamara2003} & PD \\
        & \citet{vandenbosch2006} & LD, PD \\
        & \citet{denbrok2014} & LD \\
        & \citet{lardo2015} & LD \\
        & \citet{kimmig2015} & LD \\
        
        \hline
    \end{tabular}
    
    \qquad
    
    \textbf{Notes.} Columns: (1) cluster identification in the NGC catalogue; (2) reference for literature data; (3) type of data taken from source (LD = LOS velocity-dispersion profile, LC = LOS velocity stellar catalogue, PD = PM velocity-dispersion profile).
\end{table}

Four clusters also have published PM velocity-dispersion profiles; we will later show these profiles in comparison to our own but will not use them as part of our distance estimation. All sources of literature data are shown in \autoref{table:litsource}; for each cluster we list the references for the source and the type of data taken from the source (LOS dispersion profiles, LOS stellar catalogues or PM dispersion profiles).

\begin{figure*}
    \centering
    \includegraphics[width=0.32\linewidth]{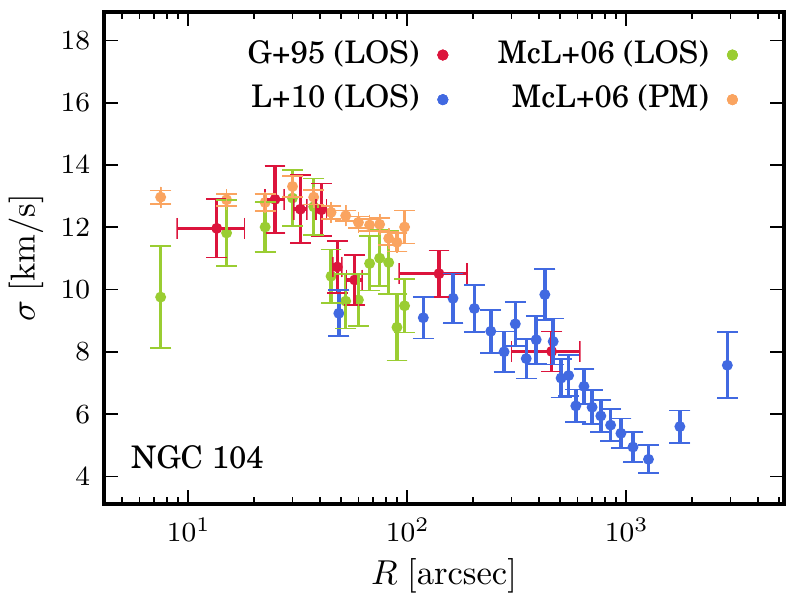}
    \quad
    \includegraphics[width=0.32\linewidth]{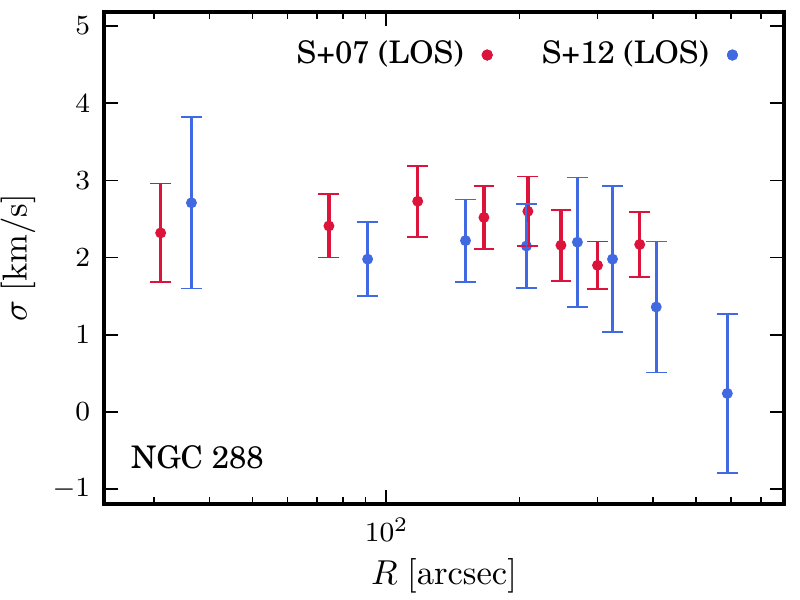}
    \quad
    \includegraphics[width=0.32\linewidth]{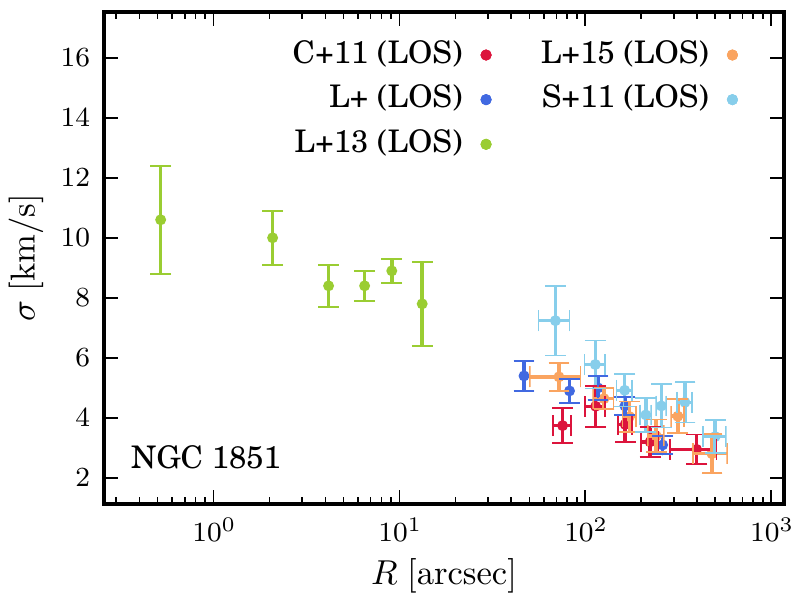}
    \includegraphics[width=0.32\linewidth]{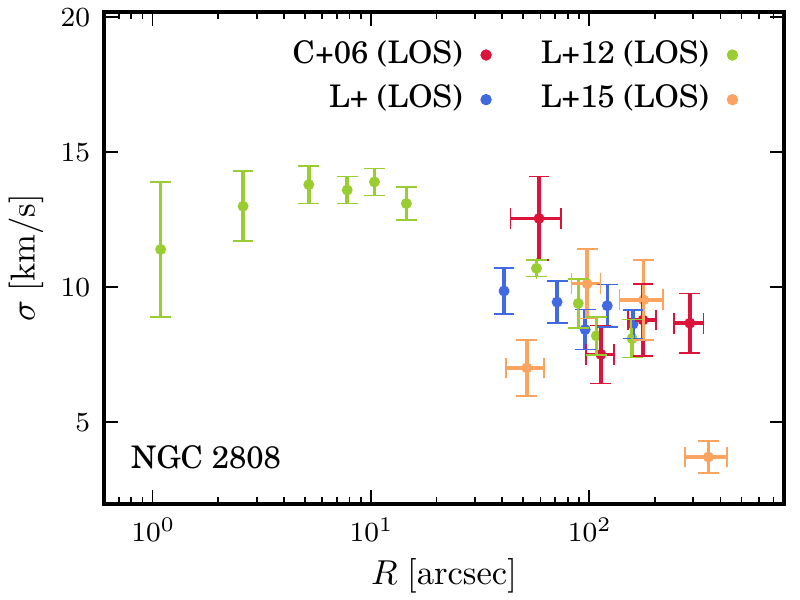}
    \quad
    \includegraphics[width=0.32\linewidth]{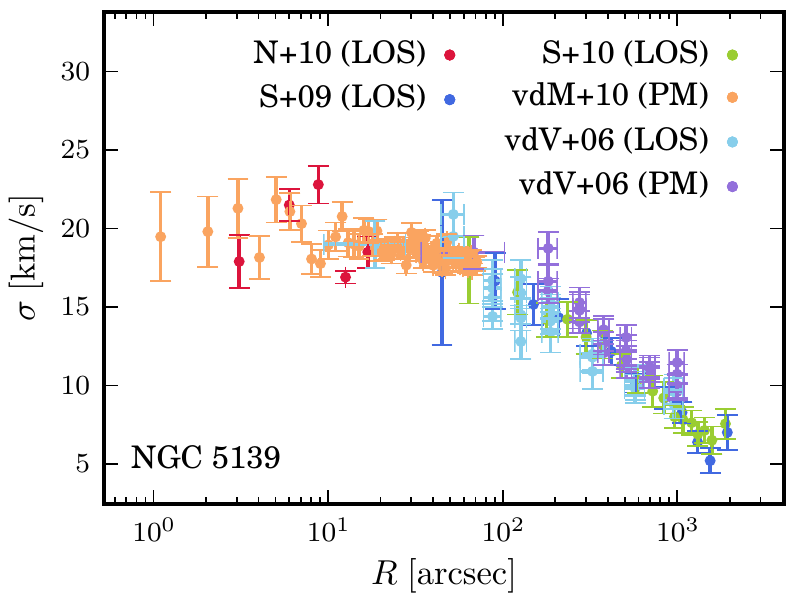}
    \quad
    \includegraphics[width=0.32\linewidth]{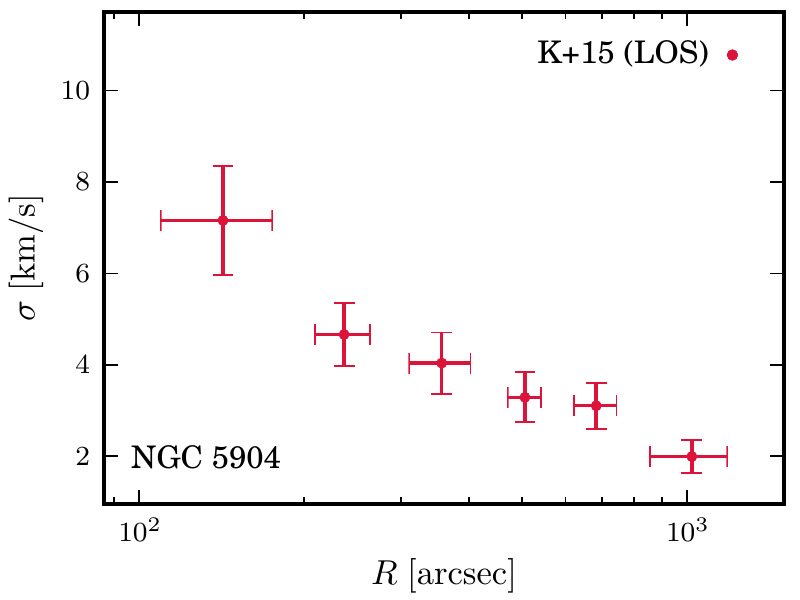}
    \includegraphics[width=0.32\linewidth]{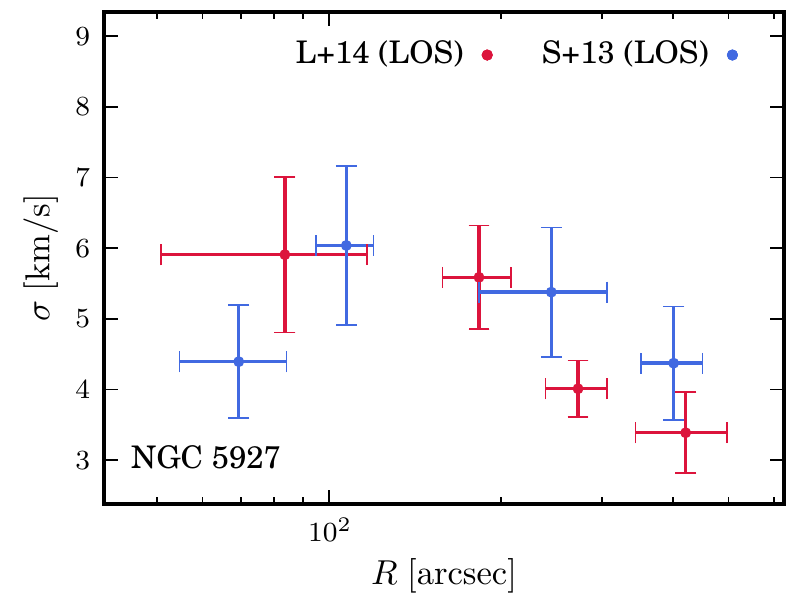}
    \quad
    \includegraphics[width=0.32\linewidth]{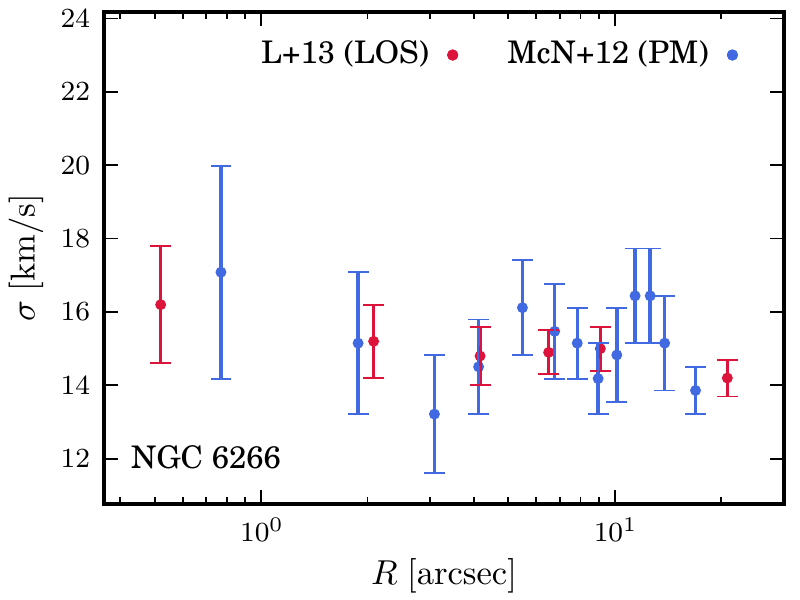}
    \quad
    \includegraphics[width=0.32\linewidth]{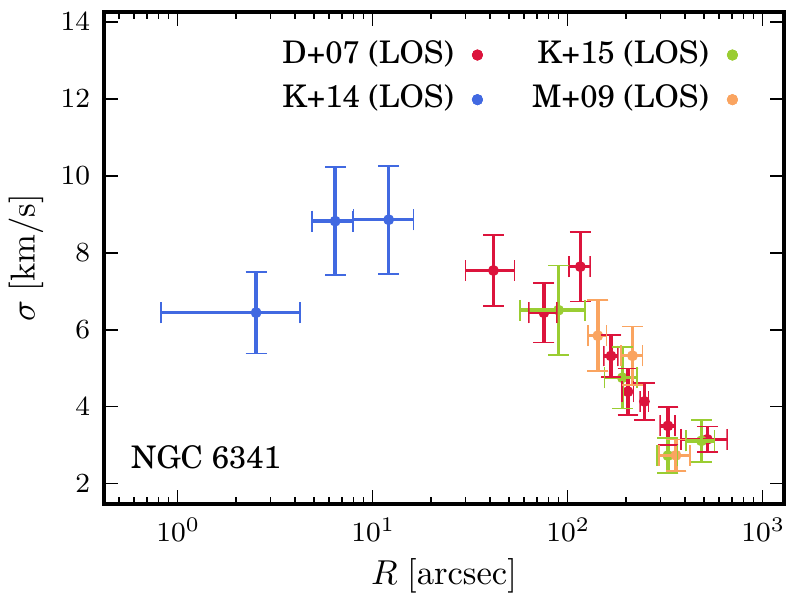}
    \includegraphics[width=0.32\linewidth]{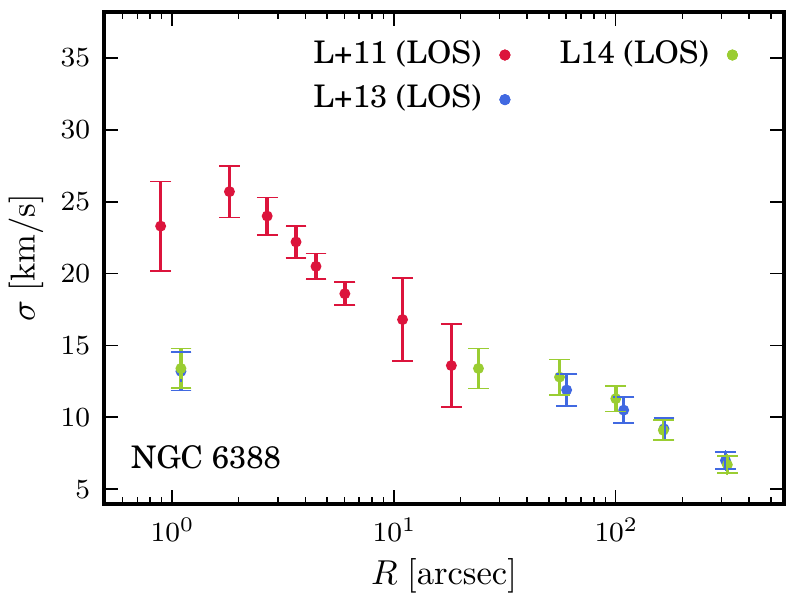}
    \quad
    \includegraphics[width=0.32\linewidth]{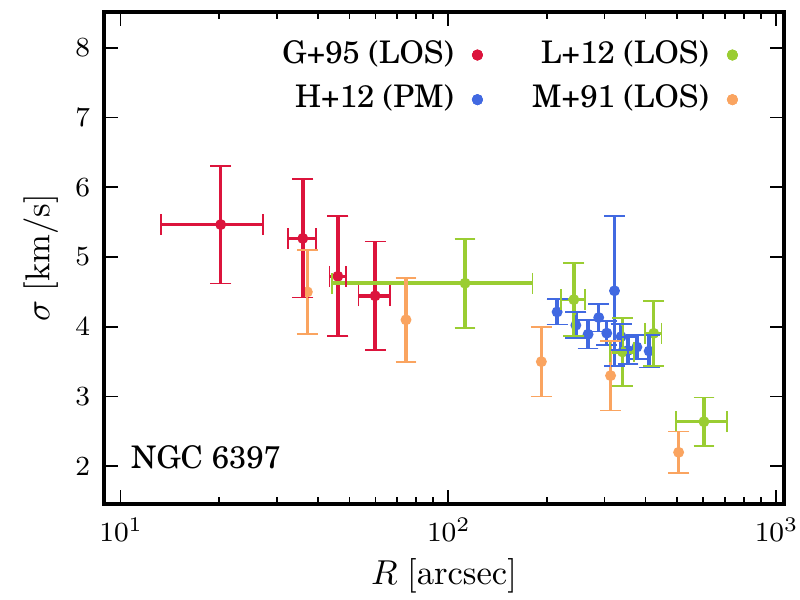}
    \quad
    \includegraphics[width=0.32\linewidth]{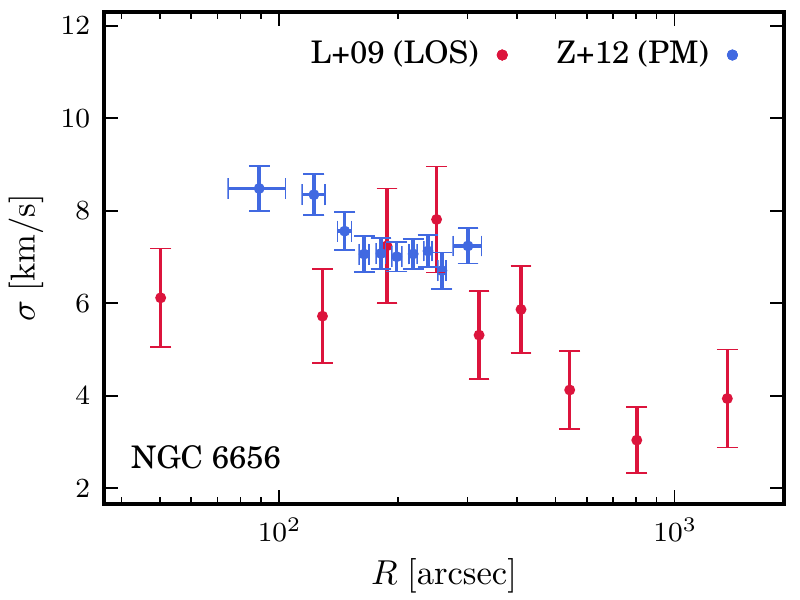}
    \includegraphics[width=0.32\linewidth]{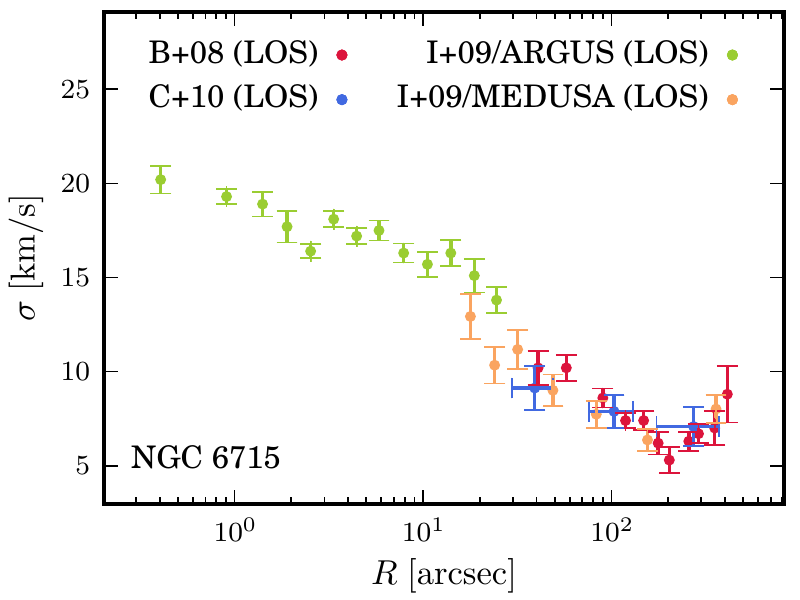}
    \quad
    \includegraphics[width=0.32\linewidth]{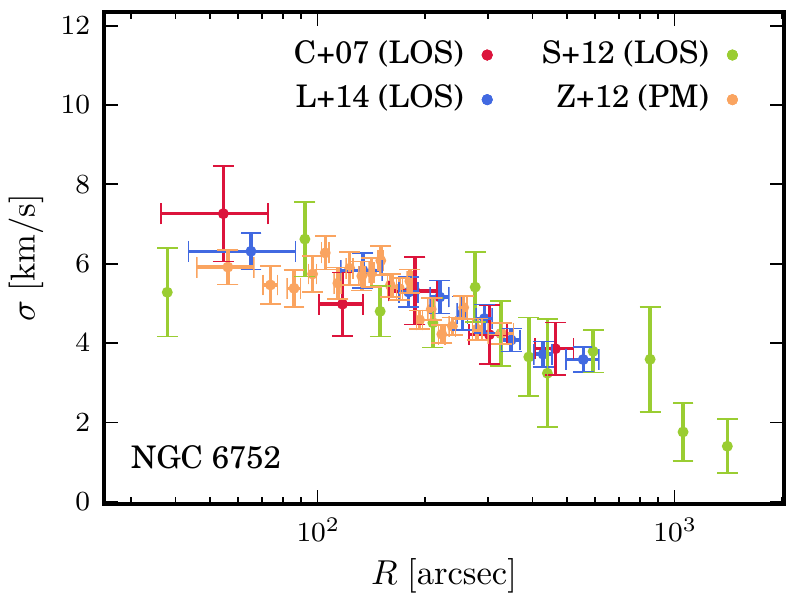}
    \quad
    \includegraphics[width=0.32\linewidth]{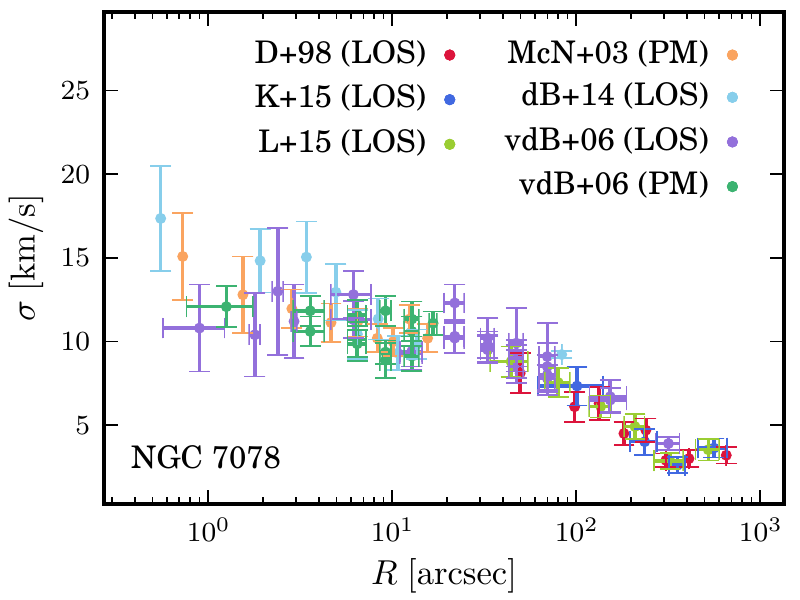}
    \caption{Compilations of literature dispersion profiles for each of our clusters. The cluster name is shown in bottom left corner of each panel. The data sources and type of data (LOS or PM) are provided in the legend. The sources are abbreviated for convenience; for the full references see \autoref{table:litsource}. PM datasets are shown at the distance provided in \citetalias{harris1996}.}
    \label{fig:literature}
\end{figure*}

We also show the compilations of literature data for each of our clusters in \autoref{fig:literature}. Cluster names are shown in the bottom-left corner of each panel. Points are coloured according to source and data type (LOS or PM), as indicated in the legends; colours are different in each panel. Source references were abbreviated, the full references are given in \autoref{table:litsource}. These compilations are also available in \autoref{app:lit}.

\subsection{MGE fits to surface-brightness profiles}
\label{sect:sbprofiles}

We begin with the Chebyshev polynomial fits to the V-band surface brightness profiles given in \citet{trager1995}. These are observed profiles and do not account for interstellar reddening. If we only cared about distances, we could proceed straight away with the uncorrected profiles, as we only need the shape of the luminosity profile for our subsequent analysis, not the normalisation factor. However, as a by-product of our analysis, we will also be able to estimate cluster mass-to-light (M/L) ratios and masses; for these M/L and mass estimates to be meaningful, we must ensure the surface brightness profiles have been properly corrected. To do this, we use extinction values $A_{V}$ from \citet{schlafly2011}.\footnote{In fact, we pulled the extinction values from the NASA/IPAC Extragalactic Database.}

Next, we convert the surface brightness values from $\rm mag / arcsec^2$ to \Lsun/pc$^2$ via
\begin{equation}
    \Sigma = \frac{1}{c^2} 10^{-0.4( \mu - M_{V,\odot} + 5 )}
\end{equation}
where we have used $\Sigma$ to denote the surface brightness in units of \Lsun/pc$^2$ and $\mu$ to denote the surface brightness in units of $\rm mag / arcsec^2$. $M_{V,\odot} = 4.83$~mag is the $V$-band absolute magnitude of the Sun and $c = \pi / 648000$~rad~arcsec$^{-1}$ is the conversion factor from arcsec to radians.

\begin{figure}
    \centering
    \includegraphics[width=0.95\linewidth]{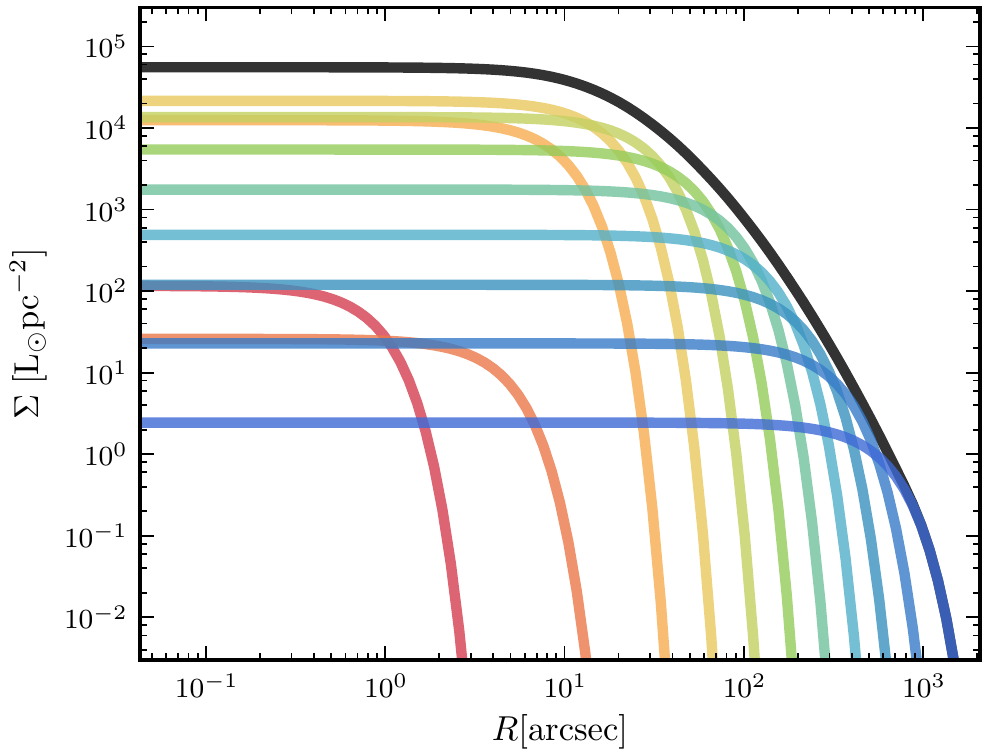}
    \caption{MGE fit to the surface-brightness profile for NGC\,2808. The black line shows the full MGE profile and the coloured lines show the individual Gaussian components.}
    \label{fig:mge_trager}
\end{figure}

Finally, to perform the MGE fitting, we use the routine described in \citet{cappellari2002}.\footnote{While this fitting approach follows the commonly accepted procedure in stellar dynamics \citep[e.g.][]{vandeven2006, vandenbosch2008, emsellem2011, luetzgendorf2012, feldmeier2013, seth2014}, we note that statistical literature suggests that alternative implementations (such as maximum likelihood estimation with expectation maximisation) may be worth exploring in the future.} We start the fitting procedure with 15-component fits, however many of the final fits contain fewer Gaussian components as the routine itself discards components that make a negligible contribution to the overall profile. As an example, \autoref{fig:mge_trager} shows the surface brightness MGE for NGC\,2808, broken down into its constituent Gaussians. The coloured lines show the individual Gaussian components and the black line shows the total MGE profile.

\begin{figure*}
    \centering
    \includegraphics[width=0.32\linewidth]{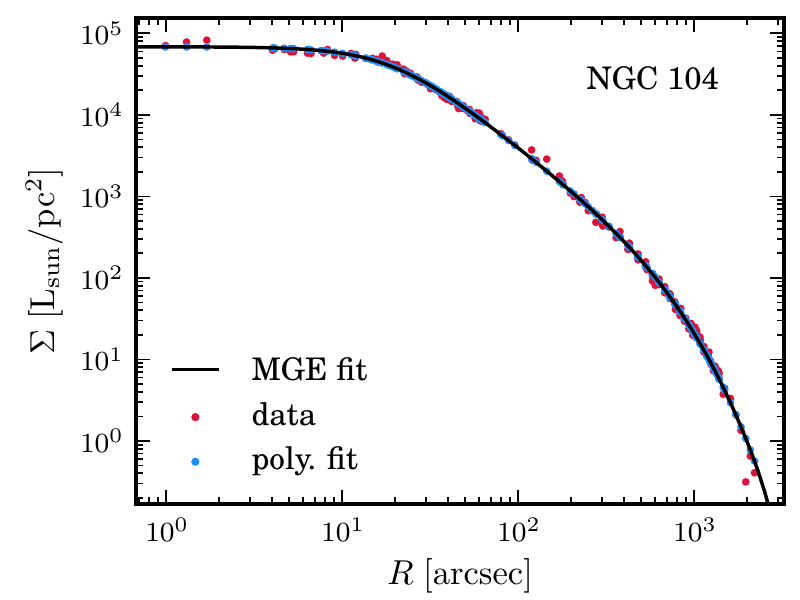}
    \quad
    \includegraphics[width=0.32\linewidth]{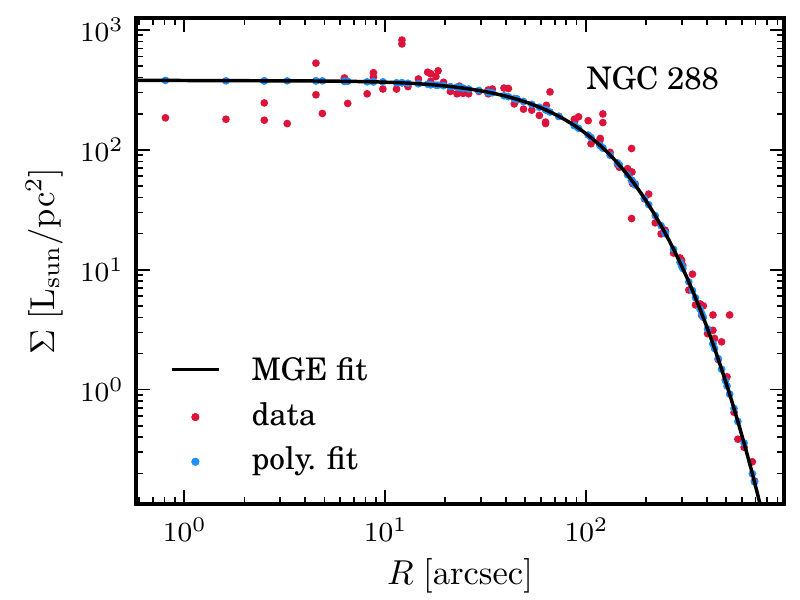}
    \quad
    \includegraphics[width=0.32\linewidth]{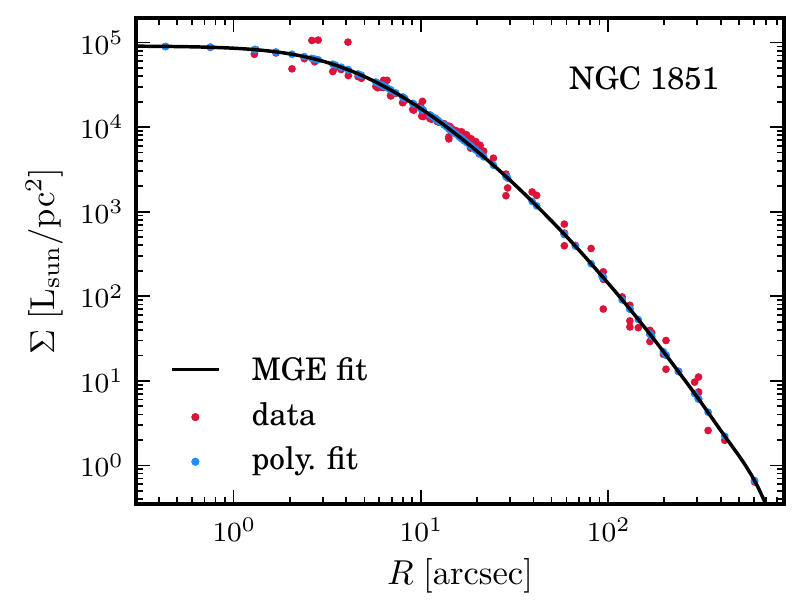}
    \includegraphics[width=0.32\linewidth]{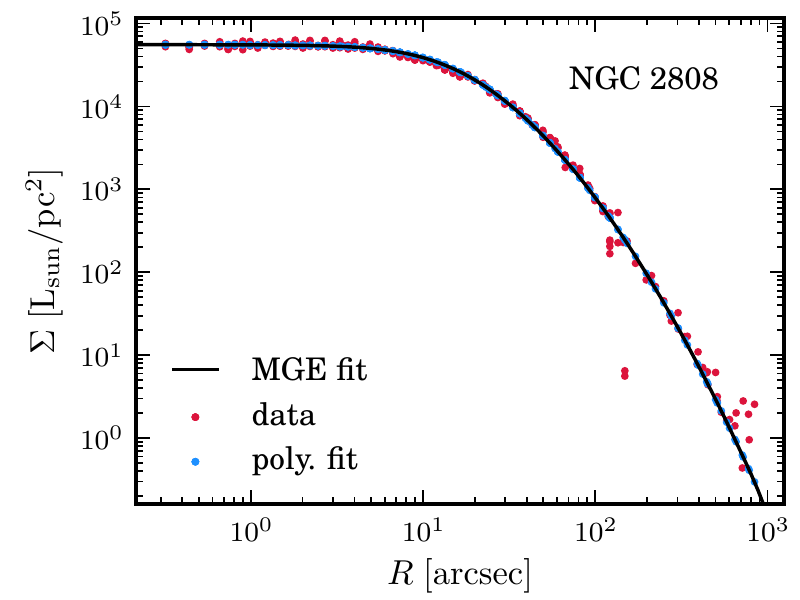}
    \quad
    \includegraphics[width=0.32\linewidth]{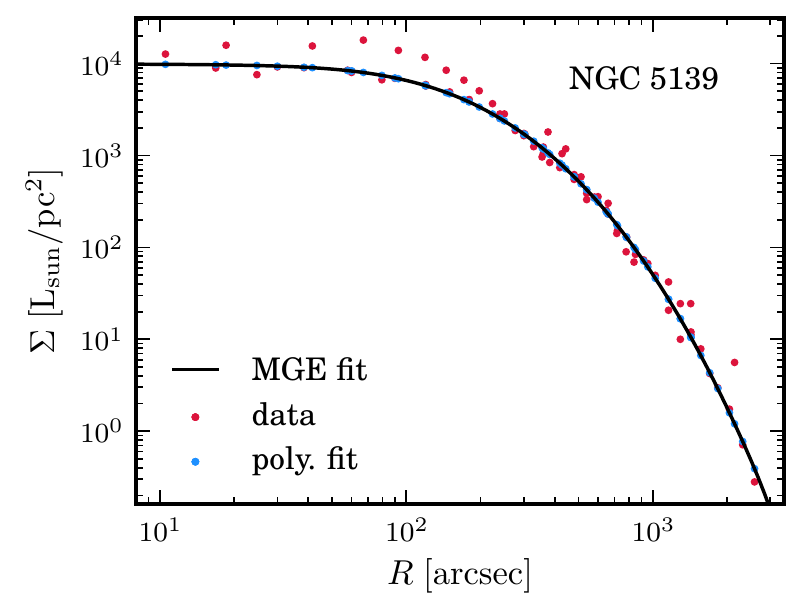}
    \quad
    \includegraphics[width=0.32\linewidth]{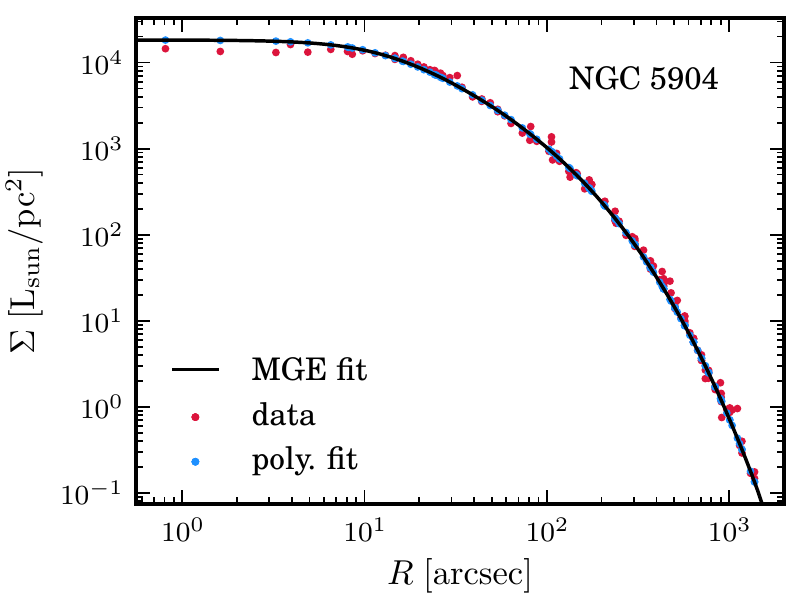}
    \includegraphics[width=0.32\linewidth]{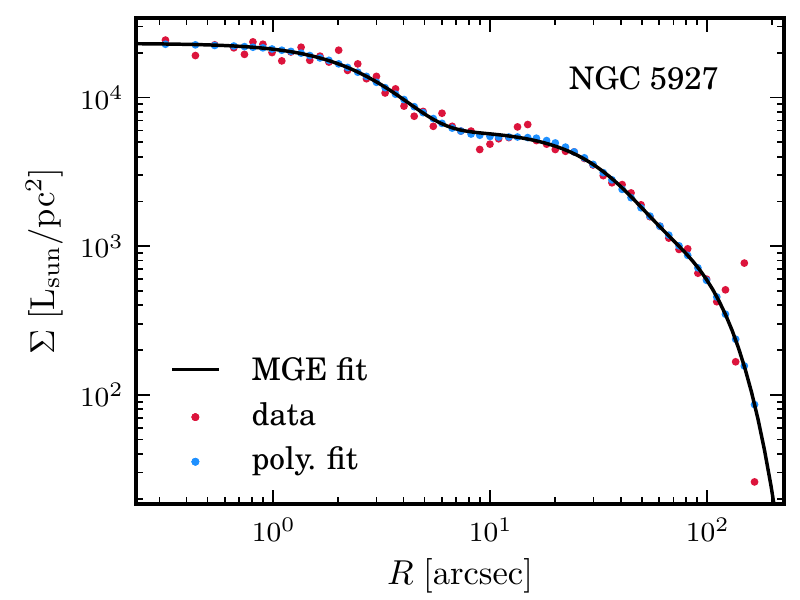}
    \quad
    \includegraphics[width=0.32\linewidth]{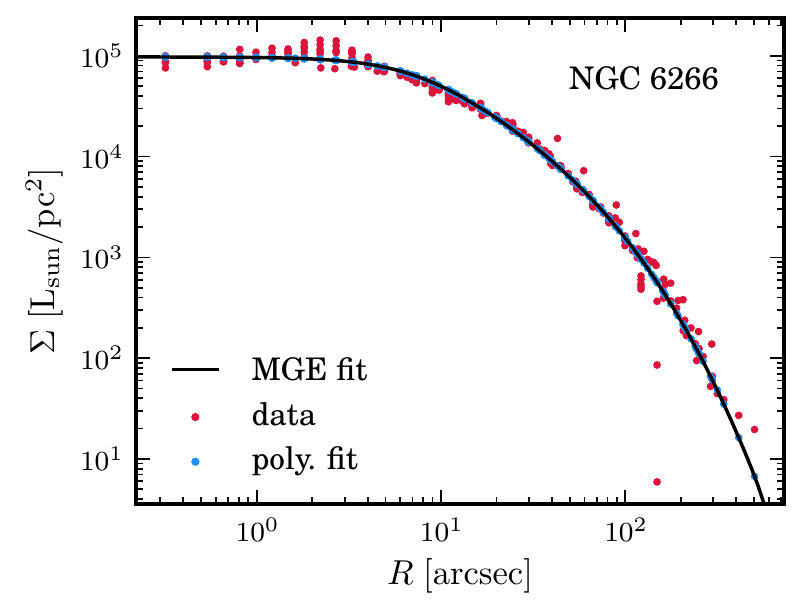}
    \quad
    \includegraphics[width=0.32\linewidth]{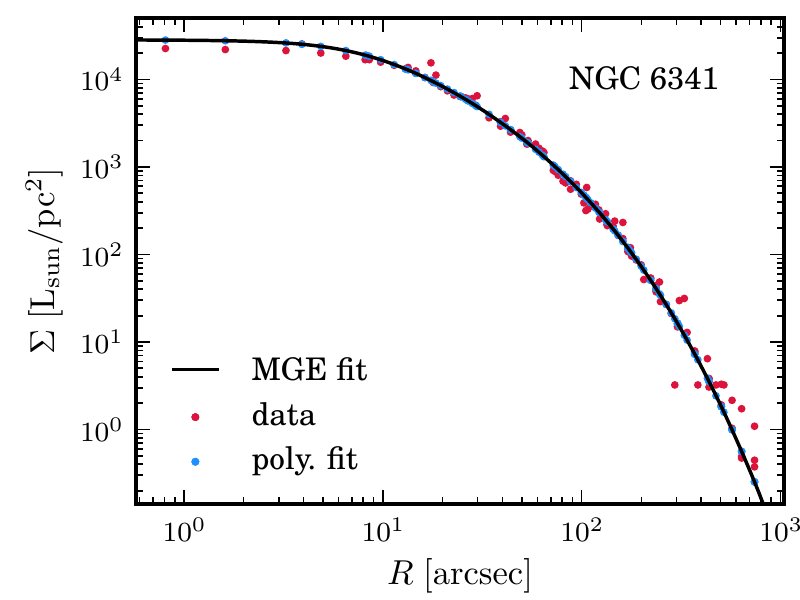}
    \includegraphics[width=0.32\linewidth]{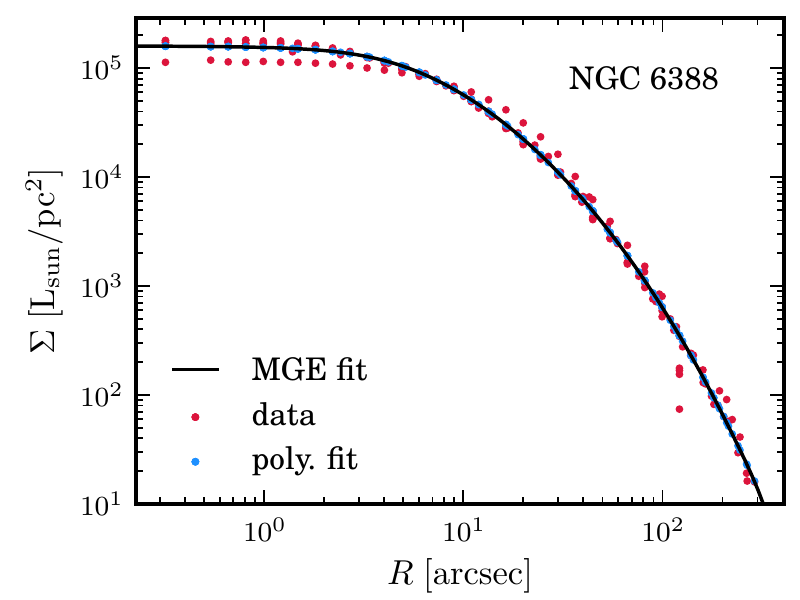}
    \quad
    \includegraphics[width=0.32\linewidth]{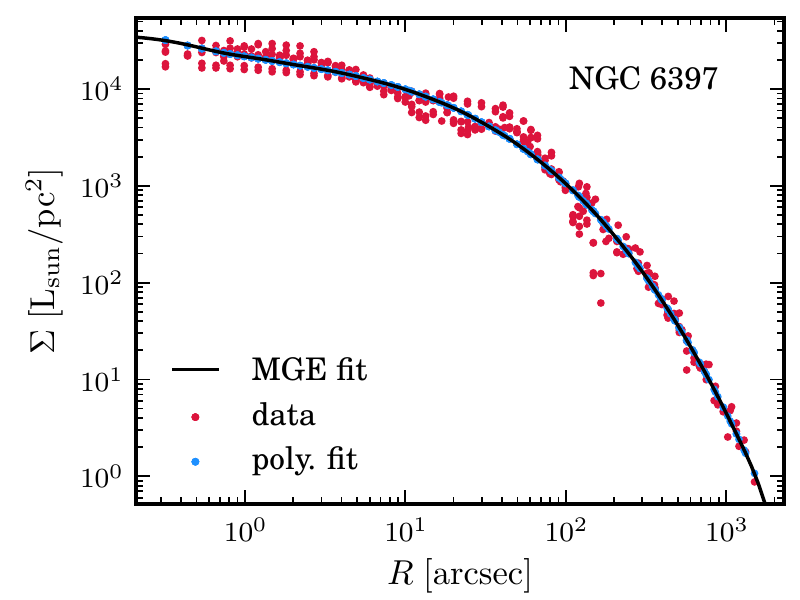}
    \quad
    \includegraphics[width=0.32\linewidth]{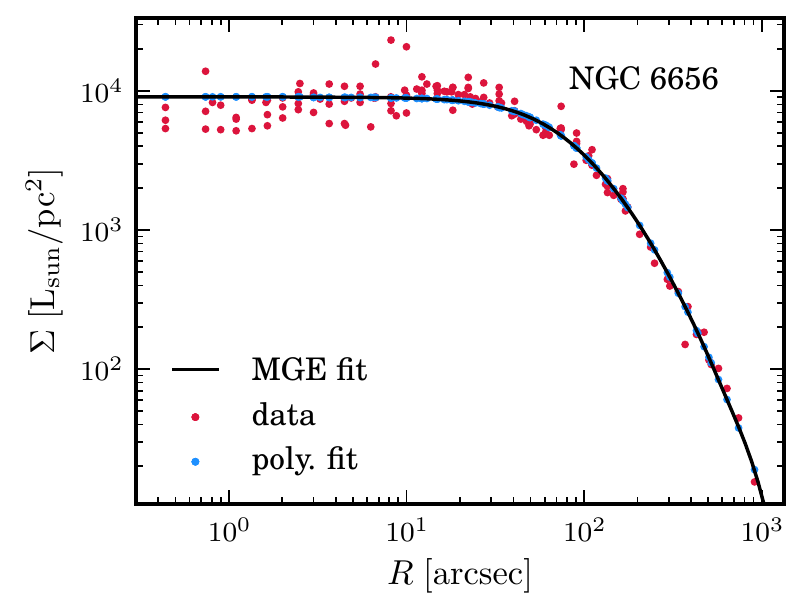}
    \includegraphics[width=0.32\linewidth]{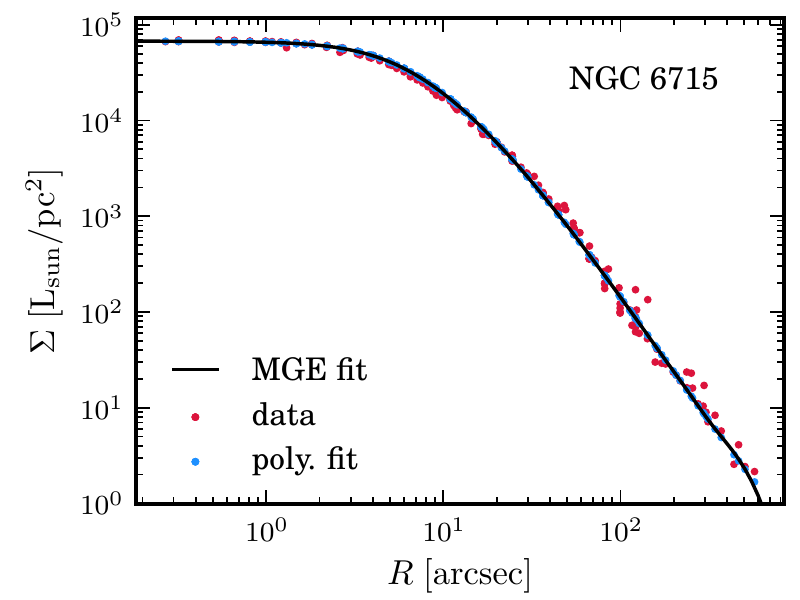}
    \quad
    \includegraphics[width=0.32\linewidth]{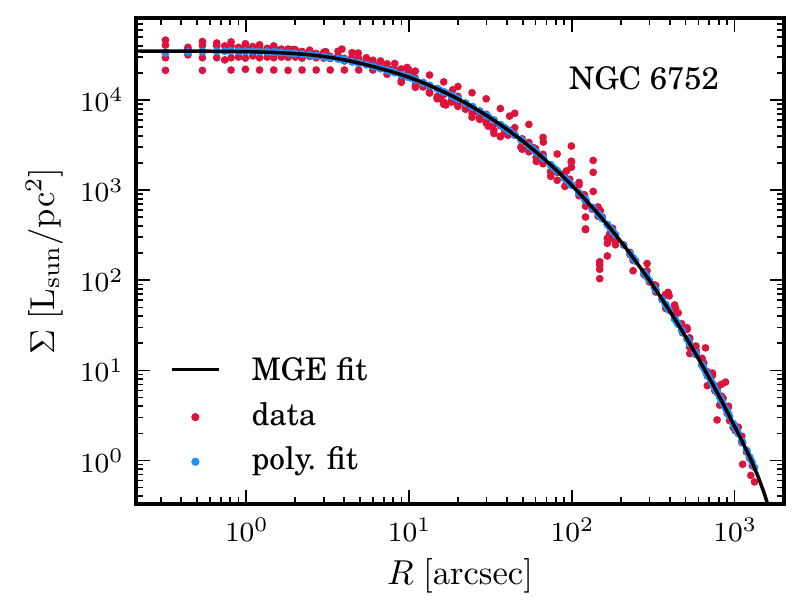}
    \quad
    \includegraphics[width=0.32\linewidth]{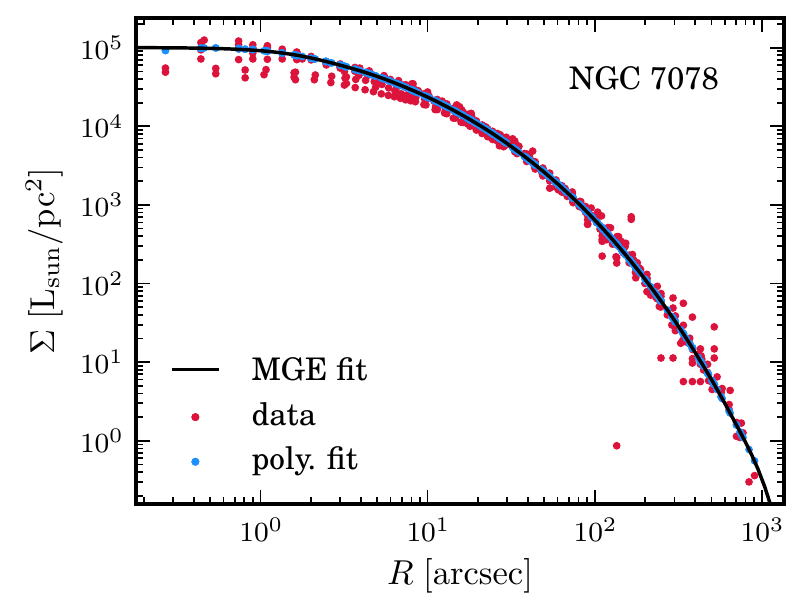}
    \caption{V-band surface brightness profiles for each cluster. The cluster name is shown in the top-right corner of each panel. Red points shows the data compiled by \citet{trager1995}. Blue points show the Chebyshev polynomial fits from \citet{trager1995}. Black lines show our MGE fits to the Chebyshev polynomials.}
    \label{fig:sbprofiles}
\end{figure*}

The V-band surface brightness profiles and our MGE fits are shown in \autoref{fig:sbprofiles}. Both sets of data points are taken from \citet{trager1995}: red points show the surface brightness data, blue points show their Chebyshev polynomial fits. Our MGE fits to the Chebyshev polynomials are shown as black lines.


\section{Distance estimation method}
\label{sect:distest}

Dynamical distance estimation relies on the fact that LOS velocities are measured as physical distance per unit time, while PMs are measured as angular distance per unit time. To convert physical distance to angular distance, or vice versa, we require the distance to the object. For an object with PM $\mu$ (in \masyr), its transverse velocity $v$ (in \kms) is given by
\begin{equation}
    v = D C \mu,
\end{equation}
where $D$ is the distance of the object (in kpc) and $C$ is a constant that takes care of the remaining unit conversions.\footnote{$C = 4.74047$~km\,yr\,kpc$^{-1}$\,mas$^{-1}$\,s$^{-1}$.} Dispersions $\sigma_{v}$ and $\sigma_{\mu}$ scale similarly. So, if we knew both $\sigma_{\mu}$ and $\sigma_v$, we could then infer the distance to the object by finding the factor by which the PM velocity dispersion profile must be scaled to best match the LOS velocity dispersion. Of course, this implicitly assumes that the object is isotropic so that, by definition, the PM and LOS dispersions are the same. If the object is anisotropic, then the PM dispersion will be some factor of the LOS dispersion, and this must be accounted for in the distance estimation. We showed in \citetalias{watkins2015} that the clusters in our sample are very close to isotropic, particularly at their centres, so we will assume that the clusters are isotropic over the range for which we have data.

So by comparing the dispersion profiles we obtain from our PM catalogues against LOS velocity dispersion profiles from the literature, we can estimate distances to our clusters. However, our PM dispersion profiles and the LOS dispersion profiles are not sampled at the same radii and, even worse, in some cases, the radial overlap is small or non-existent. This makes it challenging to determine for which distance the two profiles best agree. Therefore, instead of comparing profiles directly, we fit a simple dynamical model to both dispersion profiles and then determine for which distance the best-fit models agree. We determine the best-fitting models using maximum likelihood analysis.

Here we describe our dynamical models and then lay out the likelihood functions we use to assess the quality of the model fits.

\subsection{Models}
\label{sect:models}

We use the Jeans Axisymmetric MGE (JAM) models described in \citet{watkins2013}, which were expanded from \citet{cappellari2008}. We use only the model-predicted LOS dispersion profiles as, under the assumption of isotropy, the LOS and PM profiles are the same. The JAM models require projected luminosity-density and mass-density profiles, both in the form of an MGE. In general, these may be different, however, as we will discuss, we will assume that the mass-density profiles have the same shape as the luminosity-density profiles so that all that is required is the surface brightness profiles (already described in \autoref{sect:sbprofiles}) and a mass-to-light ratio.

These models allow for axisymmetry, anisotropy, rotation and a mass-to-light ratio that varies with radius. However, for this study, we assume that the clusters are spherical, isotropic, non-rotating and have a constant mass-to-light ratio. These assumptions are typical of dynamical cluster studies \citep[eg.][]{mclaughlin2003, sollima2010}. Although typical, it does not necessarily follow that they are correct; however, in general, we believe that these assumptions are reasonable for most of the clusters. The exception is \wcen, which we know is highly flattened, shows significant rotation and is anisotropic \citep[e.g.][]{vandeven2006, vandermarel2010, bianchini2013, watkins2013} (although we do believe the mass-to-light ratio is constant \citep{vandeven2006}). Nevertheless, we include \wcen\ in our analysis, however, we will treat the results with caution. We will discuss the model assumptions in detail in \autoref{sect:assumptions}.

\subsection{Likelihood functions}
\label{sect:likelihoods}

Let us assume that we have a LOS velocity-dispersion profile consisting of dispersions $\sigma_{\rm L,i} \pm \delta_{\rm L,i}$ (in \kms) measured at radii $R_{\rm L,i}$ (in arcsec) and a PM velocity-dispersion profile consisting of dispersions $\sigma_{\rm P,j} \pm \delta_{\rm P,j}$ (in \masyr) measured at radii $R_{\rm P,j}$ (in arcsec). The units here are important, as we will see.

Let us also assume that we have a dynamical model for which we can calculate the dispersion $\sigma_{\rm model}$ (in units of \kms) at radius $R$ for a given distance $D$ and mass-to-light ratio $\Upsilon$\footnote{As we use V-band surface brightness profiles in our dynamical models, all our M/L ratios are in the V-band.}. Under the assumptions we laid out above, both distance and mass-to-light ratio are the only free parameters in our models and both are multiplicative factors: the former because the models are typically calculated for $R$ in physical units whereas we require $R$ in angular units to compare with data; and the latter under the assumption that the mass-to-light ratio is constant through the cluster. In this case, the model dispersion is
\begin{equation}
    \sigma_{\rm model} \left( R, D, \Upsilon \right) = \sqrt{\frac{D \Upsilon}{D_0 \Upsilon_0}} \sigma_{\rm model} \left( R, D_0 \Upsilon_0 \right)
\end{equation}
for some reference distance $D_0$ and mass-to-light ratio $\Upsilon_0$. This simplifies the modelling process as we only need to calculate the model one time for $(D_0, \Upsilon_0)$ and can then scale the model appropriately.\footnote{In principle, we could use any reference values here; for simplicity, we will use reference distance $D_0 = 1$~kpc and reference mass-to-light ratio $\Upsilon = 1$~\Msun/\Lsun.} For convenience, let us define
\begin{equation}
    \sigma_0 \left( R \right) = \frac{\sigma_{\rm model} \left( R, D_0 \Upsilon_0 \right)}{\sqrt{D_0 \Upsilon_0}}.
\end{equation}

We use likelihood maximisation to determine which models best fit the data. For a particular distance and mass-to-light ratio, the log-likelihood function for fits to the LOS velocity dispersion profile is
\begin{equation}
    \ln \mathcal{L}_{\rm L} \left( D, \Upsilon \right) = \sum_i -\frac{ \left[ \sqrt{D \Upsilon} \sigma_0 \left( R_{\rm L,i} \right) - \sigma_{\rm L,i} \right]^2}{2 \delta_{\rm L,i}^2}.
    \label{eqn:logllos}
\end{equation}
With LOS velocities alone, the best we can do is determine factor
\begin{equation}
    f = \sqrt{D \Upsilon}
\end{equation}
that is required to bring the model into good agreement with the data. We cannot distinguish how much of $f$ comes from the distance scaling and how much comes from the mass-to-light scaling.

For the PM velocity dispersion profile, the log-likelihood function for a given distance and mass-to-light ratio is
\begin{equation}
    \ln \mathcal{L}_{\rm P} \left( D, \Upsilon \right) = \sum_j -\frac{ \left[ \frac{\sqrt{D \Upsilon}}{D C} b \left( R_{\rm P,j} \right) \sigma_0 \left( R_{\rm P,j} \right) - \sigma_{\rm P,j} \right]^2}{2 \delta_{\rm P,j}^2}
    \label{eqn:loglpm}
\end{equation}
where the extra factor of $D C$ converts the units of $\sigma_0$ from \kms to \masyr and $b(R)$ is an anisotropy term that controls the size of PM dispersions relative to the LOS dispersions. Under the assumption of isotropy, as we use here, $b = 1$ at all radii. Again, with PMs alone, the best we can do is determine factor
\begin{equation}
    g = \frac{\sqrt{D \Upsilon}}{D} = \sqrt{\frac{\Upsilon}{D}}
\end{equation}
but cannot distinguish the separate contributions made by distance and mass-to-light ratio.

However, with PMs and LOS velocities together, we can determine both distance $D$ and mass-to-light ratio $\Upsilon$; the total log likelihood is then
\begin{equation}
    \ln \mathcal{L} \left( D, \Upsilon \right) = \ln \mathcal{L}_{\rm P} \left( D, \Upsilon \right) + \ln \mathcal{L}_{\rm L} \left( D, \Upsilon \right).
\end{equation}
Although $D$ and $\Upsilon$ are the free parameters of interest, they are highly correlated. As $f$ and $g$ are uncorrelated, they are more effective to use in a parameter search, as we illustrate in \autoref{sect:distances}. So in practice, we seek to find factors $f$ and $g$, then we determine distance $D$ and mass-to-light ratio $\Upsilon$ via
\begin{equation}
    D = \frac{f}{g} \qquad \mathrm{and} \qquad \Upsilon = f g.
    \label{eqn:conversions}
\end{equation}


\section{Results}
\label{sect:results}

Now that we have described both the data and the methods we will use, we are ready to estimate distances and M/L ratios for our clusters. Using the surface brightness profiles and M/L estimates, we can also estimate cluster masses.

\subsection{Distance and mass-to-light ratio estimates}
\label{sect:distances}

To recap, we calculate spherical, isotropic, non-rotating JAM models for which we assume the mass-to-light ratio is constant. Under these assumptions, the model velocity-dispersion profiles have only two free parameters -- distance $D$ and mass-to-light ratio $\Upsilon$ -- and both are multiplicative factors. We fit these models to the literature LOS and our PM velocity dispersion profiles; by combining the two fits we can estimate the distance and mass-to-light ratio of the cluster.

To begin, we perform a simple likelihood maximisation test using \autoref{eqn:logllos} to approximate the best-fitting value of $f$ and then a second simple likelihood maximisation test using \autoref{eqn:loglpm} to approximate the best-fitting value of $g$. This gives us the region of parameter space where the best model is to be found, but does not give an indication of the scatter amongst the best-fitting models. To better sample the best-fit region, we then use the Markov Chain Monte Carlo (MCMC) package \textsc{emcee} developed by \citet{foremanmackey2013}, which is an implementation of the affine-invariant ensemble sampler by \citet{goodman2010}; this approach uses multiple trial points (walkers) at each step to efficiently explore the parameter space. We run our MCMC chain with 100 walkers for 100 steps, and use the last 20 steps as the final sample.

To determine the distance and mass-to-light ratio for each cluster, we first use \autoref{eqn:conversions} to convert the final MCMC sample from $(f,g)$ space to $(D,\Upsilon)$ space. Then we take the median values of each parameter as the best-fitting values and use the 15.9th and 84.1th percentile values as lower and upper uncertainty limits as these correspond to the 1-$\sigma$ confidence interval.\footnote{These uncertainties reflect scatter in the parameter estimates due to random errors; they do not account for any additional sources of scatter that could arise if the modelling assumptions were relaxed. Note that we have not attempted to constrain the shape of the dispersion profile, we have simply used the model profile as a tool to help us match the LOS and PM data and estimate distances and M/L values. As such, the quoted uncertainties reflect the agreement between the LOS and PM data, not the goodness-of-fit of the model to the data.}

\begin{figure}
    \centering
    \includegraphics[width=\linewidth]{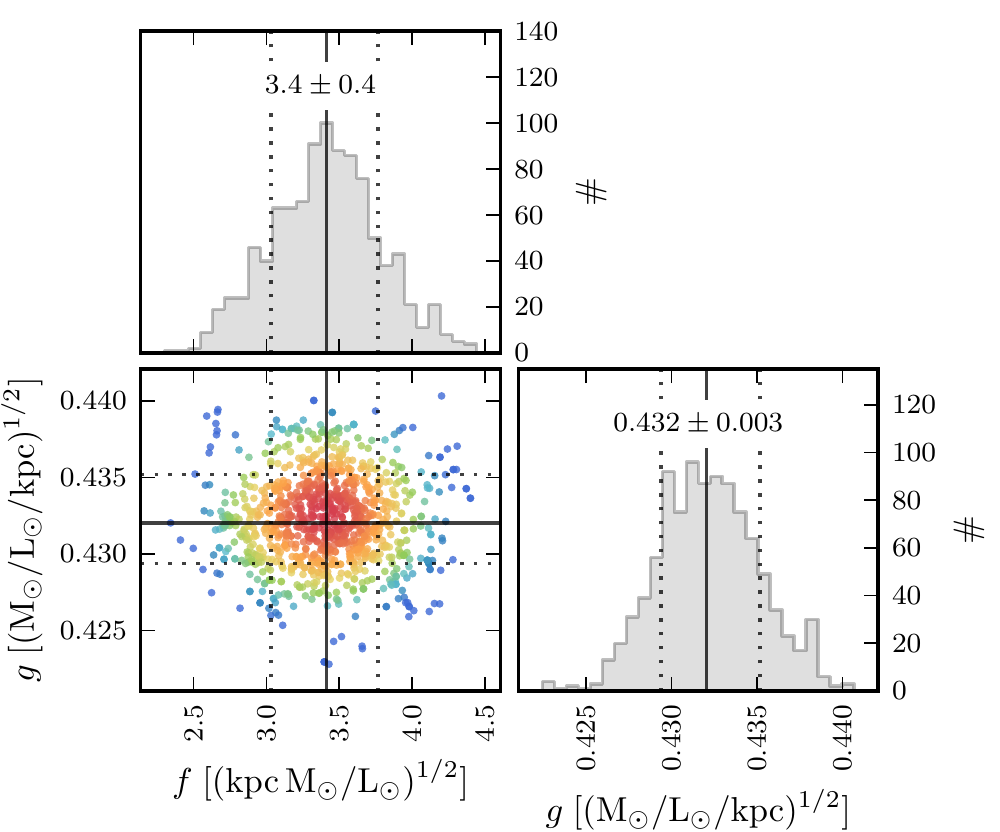}
    \includegraphics[width=\linewidth]{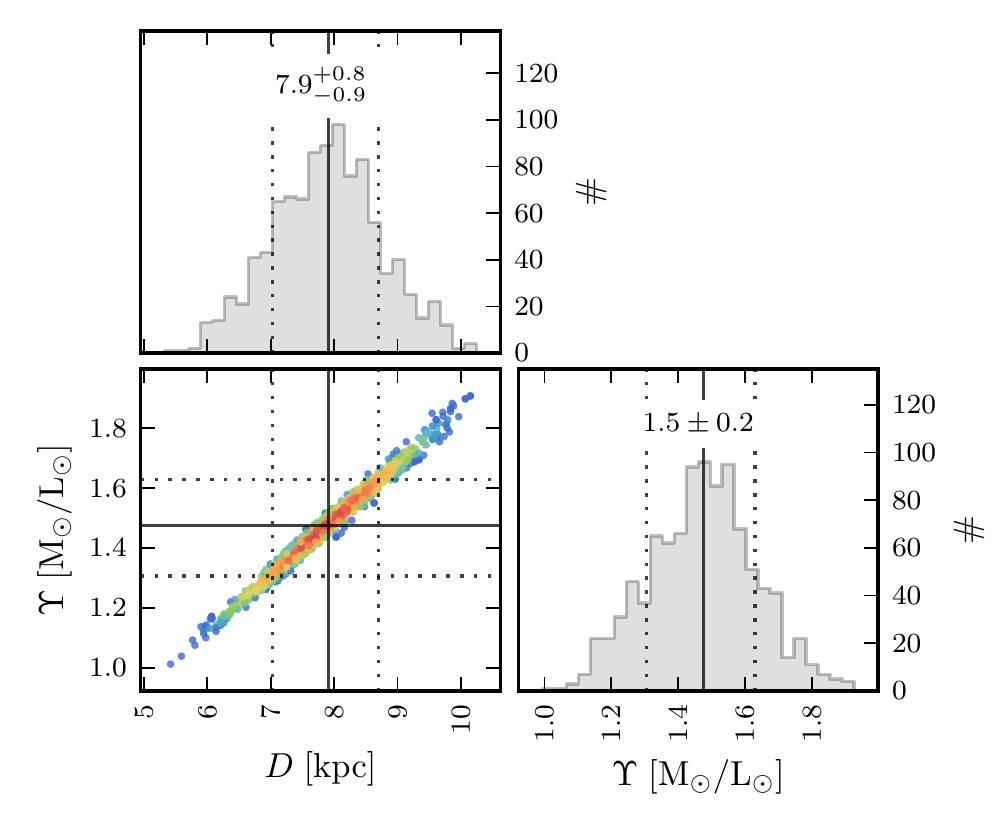}
    \caption{Final parameter distributions at the end of the MCMC run for NGC\,5927. Histograms show the individual distributions; scatter plots show the combined distributions, with points coloured by likelihood from high (red) to low (blue). The solid lines show the 50th percentiles the distributions and the dotted lines show the 15.9th and 84.1th percentiles of the distributions. In the upper panels, we show parameters $f = \sqrt{D \Upsilon}$ and $g = \sqrt{\Upsilon / D}$. In the lower panels, we show distance $D$ and mass-to-light ratio $\Upsilon$ directly. Comparing the scatter plots, it is clear that $D$ and $\Upsilon$ are strongly correlated whereas $f$ and $g$ are uncorrelated, making it more effective to perform our parameter search in $(f,g)$ space.}
    \label{fig:mcmc}
\end{figure}

As an example, \autoref{fig:mcmc} shows the final parameter distributions at the end of the MCMC run for NGC\,5927. The histograms show the distributions of the individual parameters with a best-fitting Gaussian shown as a black line. The scatter plots show the 2-dimensional distributions of the parameters, with the points coloured according to likelihood from high (red) to low (blue). The solid lines show the 50th percentiles the distributions and the dotted lines show the 15.9th and 84.1th percentiles of the distributions. The upper panels show the results for $f$ and $g$, which show no correlation. The lower panels show the results for $D$ and $\Upsilon$, highlighting their very strong correlation and illustrating why we run our MCMC chains in $(f,g)$ space.

For five of our clusters -- NGC\,104, NGC\,5139, NGC\,5927, NGC\,6388 and NGC\,6715 -- we restrict the range of the literature LOS data used for the fitting to be only those data points that overlap with our PMs. As we discuss in detail in \autoref{app:loslimits}, these cuts are an attempt to mitigate the effect of any inconsistencies outside of the range of our data.

\begin{figure*}
    \centering
    \includegraphics[width=0.32\linewidth]{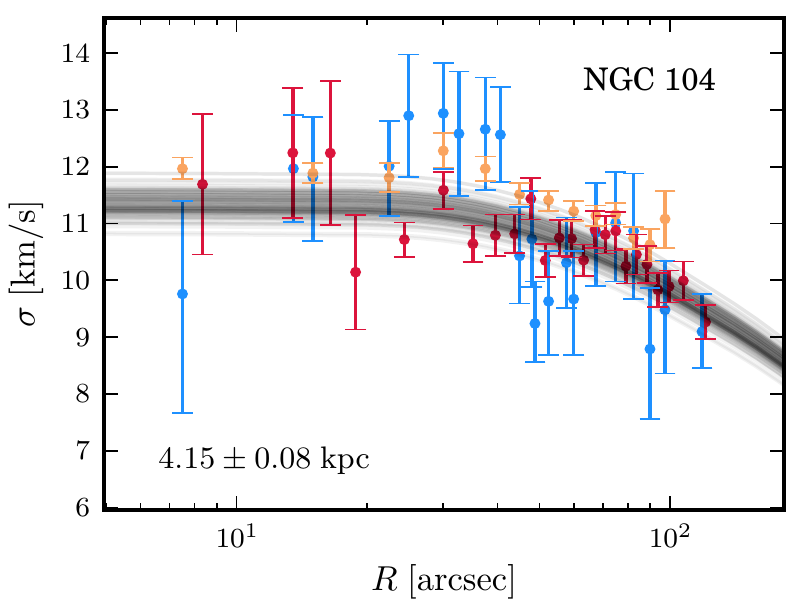}
    \quad
    \includegraphics[width=0.32\linewidth]{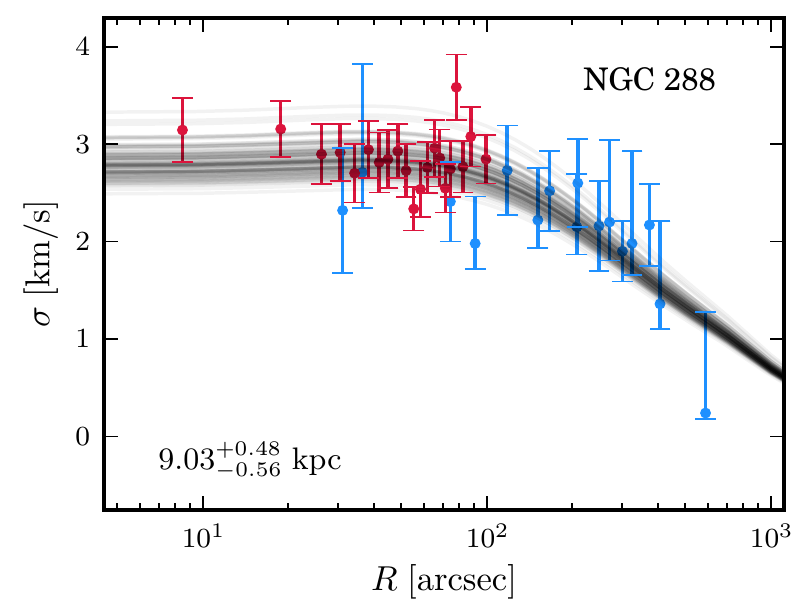}
    \quad
    \includegraphics[width=0.32\linewidth]{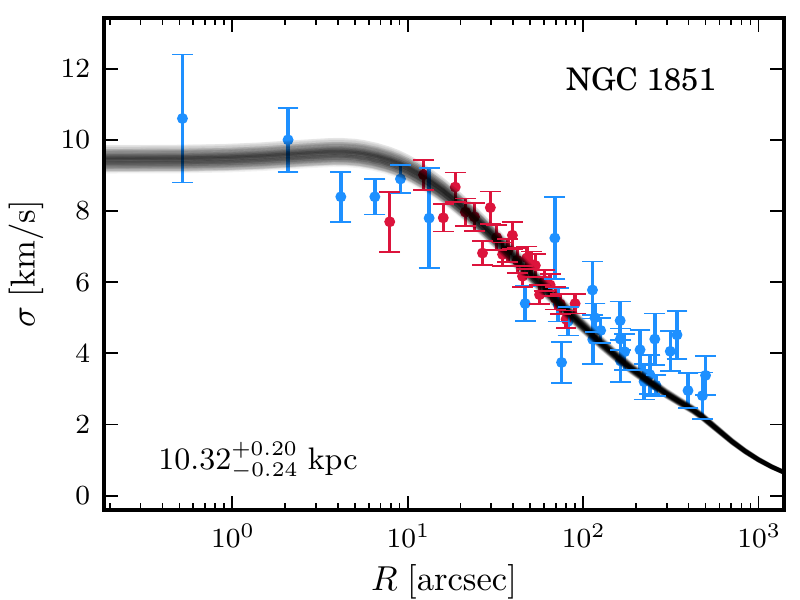}
    \includegraphics[width=0.32\linewidth]{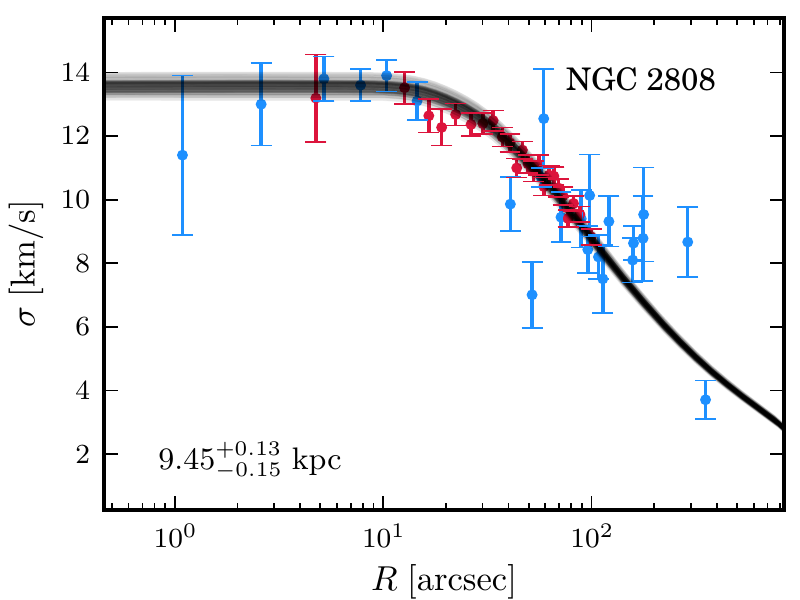}
    \quad
    \includegraphics[width=0.32\linewidth]{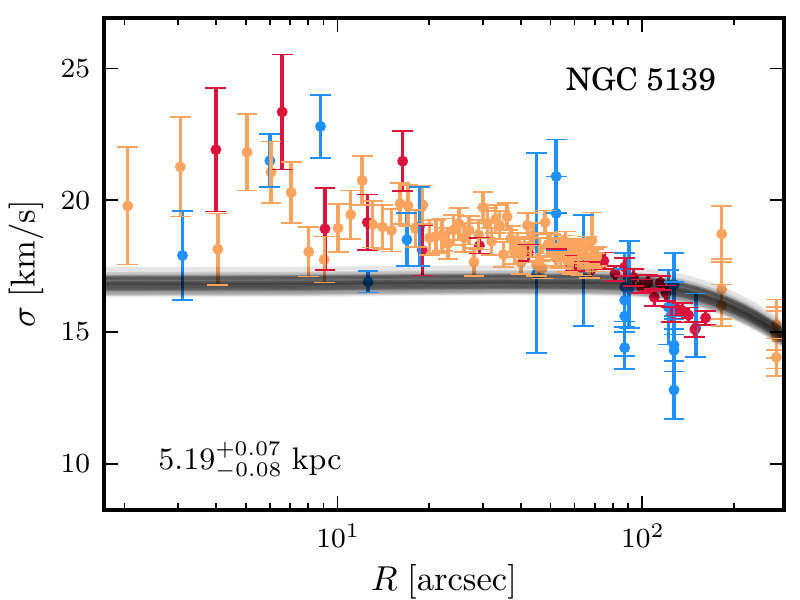}
    \quad
    \includegraphics[width=0.32\linewidth]{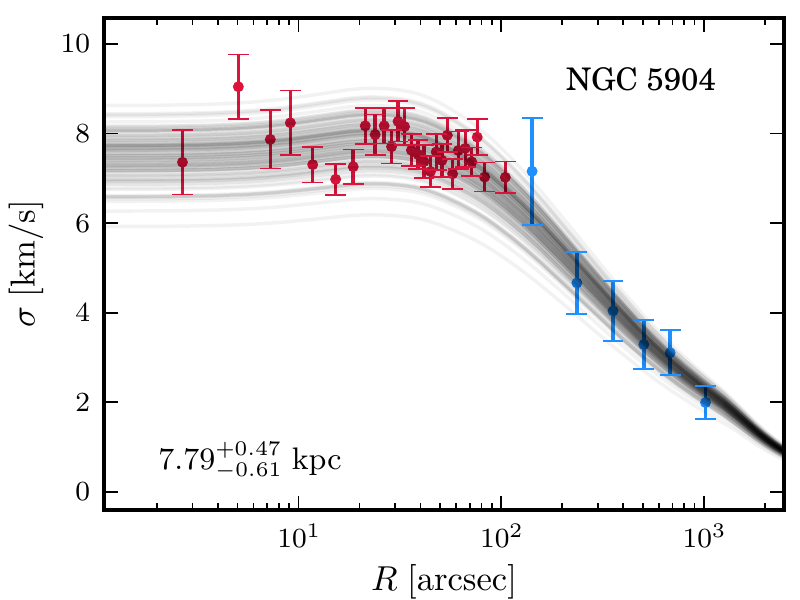}
    \includegraphics[width=0.32\linewidth]{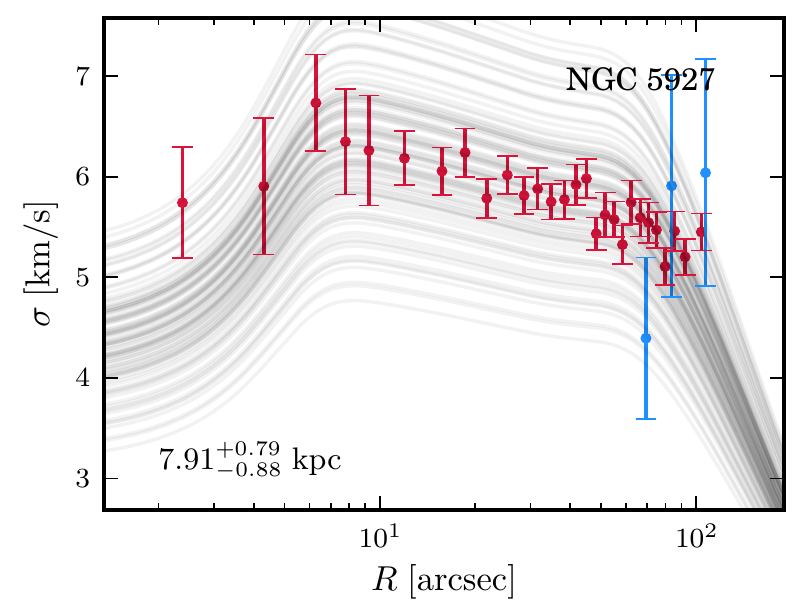}
    \quad
    \includegraphics[width=0.32\linewidth]{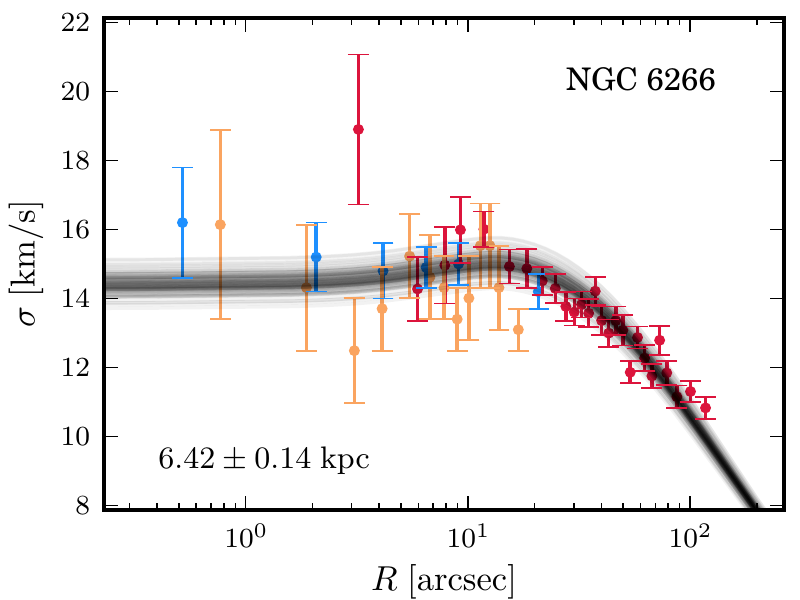}
    \quad
    \includegraphics[width=0.32\linewidth]{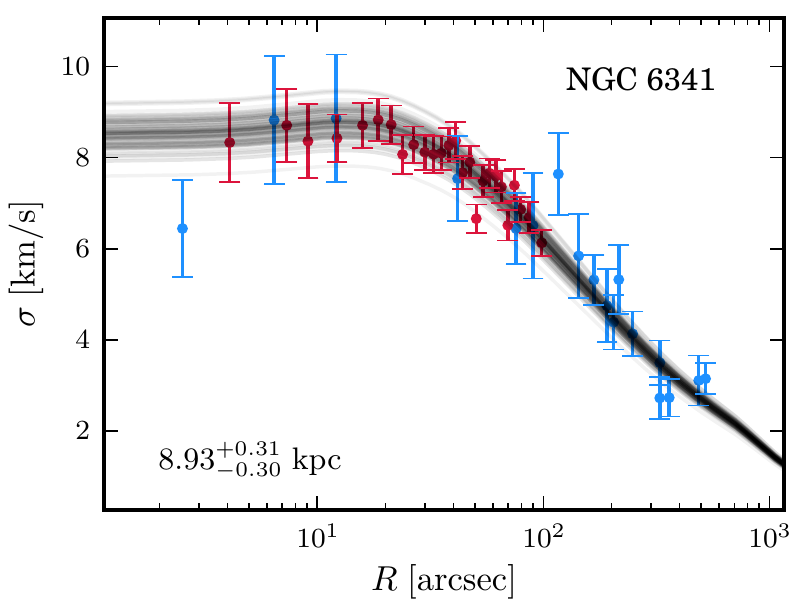}
    \includegraphics[width=0.32\linewidth]{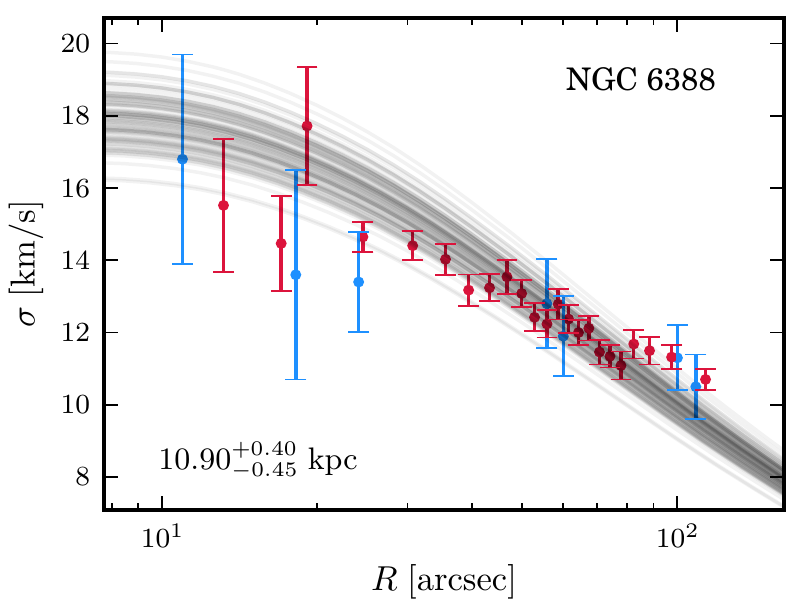}
    \quad
    \includegraphics[width=0.32\linewidth]{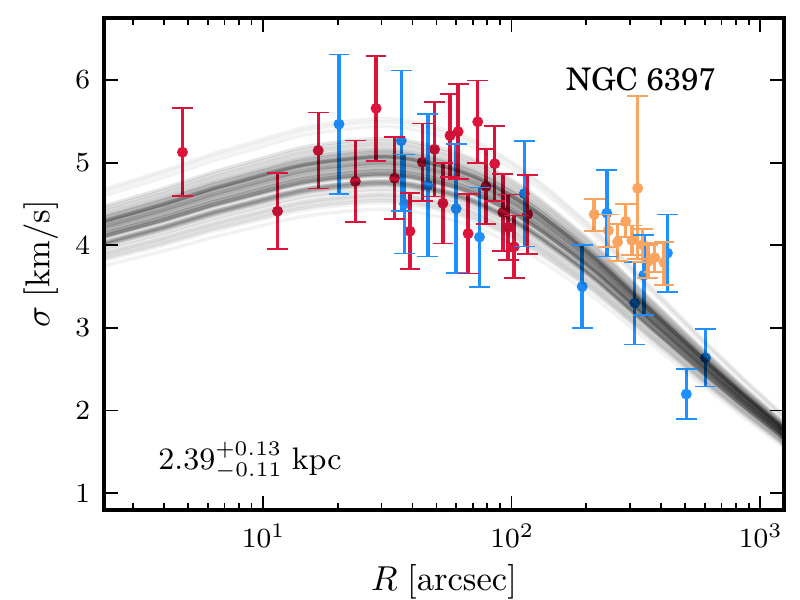}
    \quad
    \includegraphics[width=0.32\linewidth]{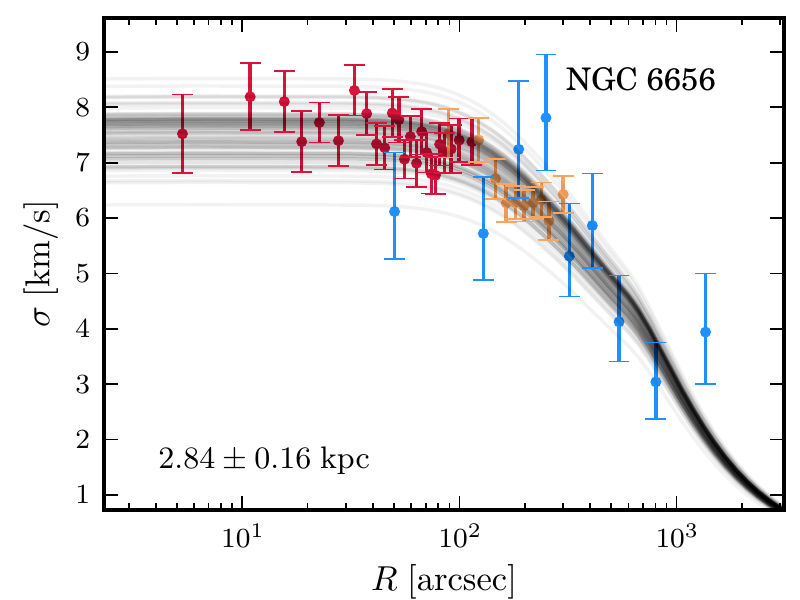}
    \includegraphics[width=0.32\linewidth]{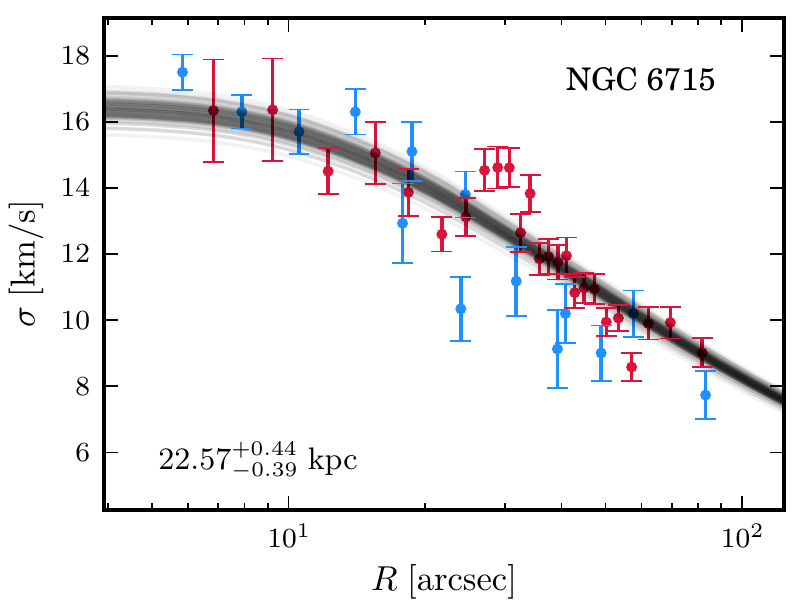}
    \quad
    \includegraphics[width=0.32\linewidth]{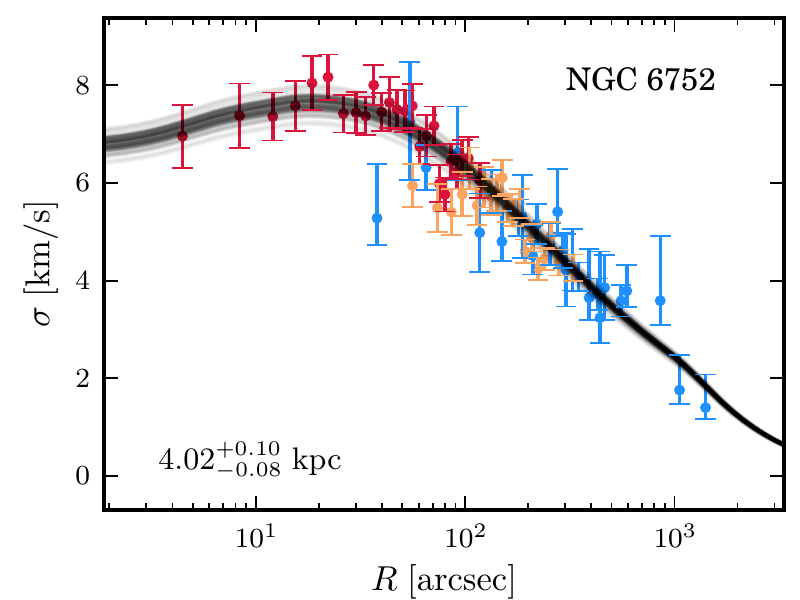}
    \quad
    \includegraphics[width=0.32\linewidth]{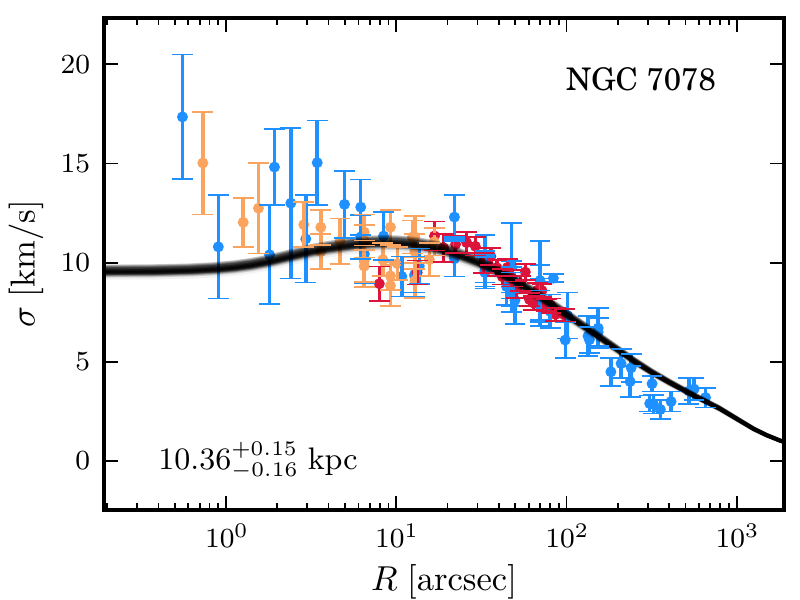}
    \caption{Velocity dispersion profiles and model fits for each of our clusters. Our PMs are shown in red, shifted to the best-fitting distance; the literature LOS velocities used for the fitting are shown in blue. 100 model draws from the final MCMC sample are shown as black lines. We give the distance estimate with uncertainties in the bottom-left corner of each panel; note that the quoted uncertainties reflect the scatter in the fit due to random errors and do not account for further systematic errors due to incorrect modelling assumptions. Where available, we show literature PM dispersion profiles (also shifted to the best-fit distance) in orange; these are shown for comparison and were not used for the fits.}
    \label{fig:results}
\end{figure*}

In \autoref{fig:results}, we show the results of our distance estimation; the cluster name is shown in the top-right corner of each panel. The blue points show the literature LOS dispersion profiles used in the fitting process. The red points show our PM dispersion profiles, shifted to the best-fitting distance. The black lines show model profiles using 100 draws from the MCMC sampling. Our distance estimate, with uncertainties, is given in the bottom-left corner of each panel. Where available, orange points show literature PM dispersion profiles shifted to the best-fitting distance; these are shown for comparison only and were not used for the fits. The best values of $f$, $g$ and the corresponding distances and mass-to-light ratios are given in \autoref{table:results}; we also give the number of PM stars that survived the cleaning and magnitude cuts in \autoref{sect:data} and were used for this analysis.

\begin{table*}
    \caption{Results: distance and mass-to-light ratio estimates.}
    \label{table:results}
    \centering
    
    \begin{tabular}{cccccccccc}
        \hline
        \hline
        Cluster & Other names & $N$ & $f$ & $g$ & $D$ & $\Upsilon_{\rm V}$ & $M_{\rm tot}$ & $D_{\rm literature}$ & $\Upsilon_{\rm V,literature}$ \\
        &  &  & (kpc\,\Msun/\Lsun)$^{0.5}$ & (\Msun/\Lsun/kpc)$^{0.5}$ & (kpc) & (\Msun/\Lsun) & (10$^5$ \Msun) & (kpc) & (\Msun/\Lsun) \\
        (1) & (2) & (3) & (4) & (5) & (6) & (7) & (8) & (9) & (10) \\
        \hline \\[-1em]
        NGC\,104  & 47\,Tuc & 5631 & $ 2.41_{-0.04}^{+0.05}$ & $0.581_{-0.004}^{+0.003}$ & $ 4.15 \pm 0.08$      & $ 1.40 \pm 0.03$      & $ 5.57_{-0.28}^{+0.33}$ &  4.5 & 2.35 \\[0.4em]
        NGC\,288  & \dots   &  713 & $ 4.44_{-0.21}^{+0.24}$ & $0.495_{-0.012}^{+0.011}$ & $ 9.03_{-0.56}^{+0.48}$ & $ 2.20_{-0.10}^{+0.13}$ & $ 0.79_{-0.11}^{+0.13}$ &  8.9 & 1.97 \\[0.4em]
        NGC\,1851 & \dots   & 2810 & $ 3.95 \pm 0.08$      & $0.383 \pm 0.004$      & $10.32_{-0.24}^{+0.20}$ & $ 1.51 \pm 0.03$      & $ 1.78_{-0.11}^{+0.10}$ & 12.1 & 1.98 \\[0.4em]
        NGC\,2808 & \dots   & 7867 & $ 3.84_{-0.06}^{+0.05}$ & $0.406 \pm 0.002$      & $ 9.45_{-0.15}^{+0.13}$ & $ 1.56 \pm 0.02$      & $ 5.91_{-0.25}^{+0.22}$ &  9.6 & 2.02 \\[0.4em]
        NGC\,5139 & \wcen   & 16519 & $ 3.72 \pm 0.05$      & $0.716 \pm 0.003$      & $ 5.19_{-0.08}^{+0.07}$ & $ 2.66 \pm 0.04$      & $34.52_{-1.43}^{+1.45}$ &  5.2 & 1.87 \\[0.4em]
        NGC\,5904 & M\,5    & 2436 & $ 3.33_{-0.25}^{+0.22}$ & $0.429 \pm 0.004$      & $ 7.79_{-0.61}^{+0.47}$ & $ 1.43_{-0.10}^{+0.09}$ & $ 3.65 \pm 0.75$      &  7.5 & 1.96 \\[0.4em]
        NGC\,5927 & \dots   & 5801 & $ 3.41_{-0.38}^{+0.36}$ & $0.432 \pm 0.003$      & $ 7.91_{-0.88}^{+0.79}$ & $ 1.48_{-0.17}^{+0.15}$ & $ 1.44_{-0.43}^{+0.49}$ &  7.7 & 2.93 \\[0.4em]
        NGC\,6266 & M\,62   & 7102 & $ 3.77_{-0.07}^{+0.08}$ & $0.587 \pm 0.004$      & $ 6.42 \pm 0.14$      & $ 2.22 \pm 0.04$      & $ 6.09_{-0.33}^{+0.39}$ &  6.8 & 1.95 \\[0.4em]
        NGC\,6341 & M\,92   & 2729 & $ 3.80_{-0.11}^{+0.13}$ & $0.426_{-0.005}^{+0.004}$ & $ 8.93_{-0.30}^{+0.31}$ & $ 1.62_{-0.05}^{+0.06}$ & $ 2.85_{-0.25}^{+0.30}$ &  8.3 & 1.93 \\[0.4em]
        NGC\,6388 & \dots   & 6746 & $ 4.28_{-0.17}^{+0.15}$ & $0.392_{-0.002}^{+0.003}$ & $10.90_{-0.45}^{+0.40}$ & $ 1.68_{-0.07}^{+0.06}$ & $ 8.27_{-0.95}^{+0.89}$ &  9.9 & 2.55 \\[0.4em]
        NGC\,6397 & \dots   &  517 & $ 2.31 \pm 0.09$      & $0.967_{-0.023}^{+0.022}$ & $ 2.39_{-0.11}^{+0.13}$ & $ 2.23_{-0.09}^{+0.10}$ & $ 0.70_{-0.08}^{+0.09}$ &  2.3 & 1.88 \\[0.4em]
        NGC\,6656 & M\,22   & 1924 & $ 2.31_{-0.12}^{+0.13}$ & $0.816 \pm 0.010$      & $ 2.84 \pm 0.16$      & $ 1.88_{-0.10}^{+0.12}$ & $ 2.49_{-0.37}^{+0.44}$ &  3.2 & 1.87 \\[0.4em]
        NGC\,6715 & M\,54   & 3790 & $ 6.61 \pm 0.10$      & $0.293_{-0.003}^{+0.002}$ & $22.57_{-0.39}^{+0.44}$ & $ 1.94 \pm 0.03$      & $11.83_{-0.53}^{+0.62}$ & 26.5 & 1.88 \\[0.4em]
        NGC\,6752 & \dots   & 1860 & $ 2.93_{-0.07}^{+0.06}$ & $0.728 \pm 0.009$      & $ 4.02_{-0.08}^{+0.10}$ & $ 2.14_{-0.06}^{+0.05}$ & $ 1.82 \pm 0.12$      &  4.0 & 1.88 \\[0.4em]
        NGC\,7078 & M\,15   & 3107 & $ 3.92 \pm 0.05$      & $0.379 \pm 0.003$      & $10.36_{-0.16}^{+0.15}$ & $ 1.49 \pm 0.02$      & $ 4.95 \pm 0.19$      & 10.4 & 1.93 \\[0.4em]
        \hline
    \end{tabular}
    
    \qquad
    
    \textbf{Notes.} Columns: (1) cluster identification in the NGC catalogue; (2) alternate name for cluster; (3) number of PM stars used for the distance estimation; (4) $f$ parameter used in fitting procedure; (5) $g$ parameter used in fitting procedure; (6) heliocentric distance estimate; (7) V-band mass-to-light ratio estimate; (8) total cluster mass estimate; (9) heliocentric distance from \citetalias{harris1996}; (10) V-band mass-to-light ratio from \citet{mclaughlin2005}.
\end{table*}

The model parameters we estimate here provide the best fit to the data under a particular set of assumptions, but this does not guarantee that the best-fitting model is actually a good fit to the data. A visual inspection of the model fits in \autoref{fig:results} can give an indication of how well the models have performed. There are three main points to consider: the fit of the model to the PM dispersion profile, the fit of the model to the LOS dispersion profile, and the agreement of the PM and LOS dispersions.

In particular, we are interested in the shape of the model dispersion profile compared with the shapes of the data dispersion profiles. The only freedom we allow here is to change the distance or the (constant) mass-to-light ratio, these parameters have no power to change the shape of the model profile. Clusters for which the model and data profiles shapes are not consistent may indicate clusters for which the assumptions we introduced in \autoref{sect:models} break down, as we will discuss further in \autoref{sect:assumptions}.

For the majority of the clusters in our sample -- NGC\,104, NGC\,288, NGC\,1851, NGC\, 2808, NGC\,5904, NGC\,6266, NGC\,6341, NGC\,6397, NGC\,6656, NGC\,6715, NGC\,6752 and NGC\,7078 -- the agreement between PM dispersions, LOS dispersions and model fits appears to be good, that is, everything is broadly consistent (albeit with no radial overlap between PM dispersions and LOS dispersions for NGC\,5904). Where available for these clusters, the literature PM data also appears to be in good agreement with the rest of the data and the models, except for NGC\,6397 where the literature PMs sit above both LOS data and model at the best-fit distance.

For the three remaining clusters, the agreements between PM data, LOS data and model are less convincing. For both NGC\,5139 (\wcen) and NGC\,6388, the shape of the model profile is a poor fit to both the PM and LOS profiles, implying that one or more of our assumptions were not valid for these clusters. Nevertheless, despite the poor models, the data profiles are consistent with each other so the distance estimation seems to have worked successfully and we include them in our subsequent analysis.

For NGC\,5927, there is considerable scatter in the best-fitting models, though the models do seem to trace our PM data fairly well. Unfortunately, the restricted radial range of the LOS data, leaves us with only 3 data points to use for that aspect of the fit. Although not ideal, this is enough for us to estimate a distance. Given the good agreement between PM data and model here, we also include this cluster in our subsequent analysis.

Several of the cluster models in \autoref{fig:results} show central dips in their dispersion profiles. This is not unexpected. Spherical isotropic models with steep density profiles generically have this feature \citep[eg.][]{binney1982}, and such dips are also seen in many dwarf elliptical galaxies \citep[eg.][]{geha2003}. Although, if GCs do indeed host IMBHs at their centres then our models would be underestimates of the dispersion near the centre and correctly accounting for the presence of an IMBH could make the dips disappear.

\subsection{Mass estimates}
\label{sect:masses}

Our primary goal here is to estimate cluster distances. However, as a by-product of the distance estimation process, we estimate the mass-to-light ratio for each cluster. As the surface brightness profiles for the clusters are also known, we can use these mass-to-light ratios to get the surface mass-density profiles for the clusters and, from there, total mass estimates for the clusters. The mass estimates are also shown in \autoref{table:results}.


\section{Discussion}
\label{sect:discussion}

We wish to compare our dynamical distance and M/L estimates against previous photometric distances and population-synthesis M/L estimates to check for biases and systematic offsets. We also consider here the effects of the assumptions we made in \autoref{sect:models}.

\subsection{Comparison with photometric distance estimates}
\label{sect:dist_comp}

We have greatly expanded the number of dynamical distance estimates available, and all have been estimated using the same method. Both of these points will enable us to make a meaningful comparison to photometric distance estimates.

Here, we compare our dynamical-distance estimates with the (photometric) estimates from \citetalias{harris1996}, which we include in \autoref{table:results}. These estimates are suitable for this comparison as they were all derived in the same way (albeit from different datasets). The estimates were made using the mean magnitude of the observed horizontal branch $V_{\rm HB}$ for each cluster. A simple calibration relation provides the absolute magnitude of the horizontal branch $M_{\rm V,HB}$ given the cluster metallicity, from which the distance modulus $\mu = V_{\rm HB} - M_{\rm V,HB}$ and, hence, distance follow.

\begin{figure}
    \centering
    \includegraphics[width=\linewidth]{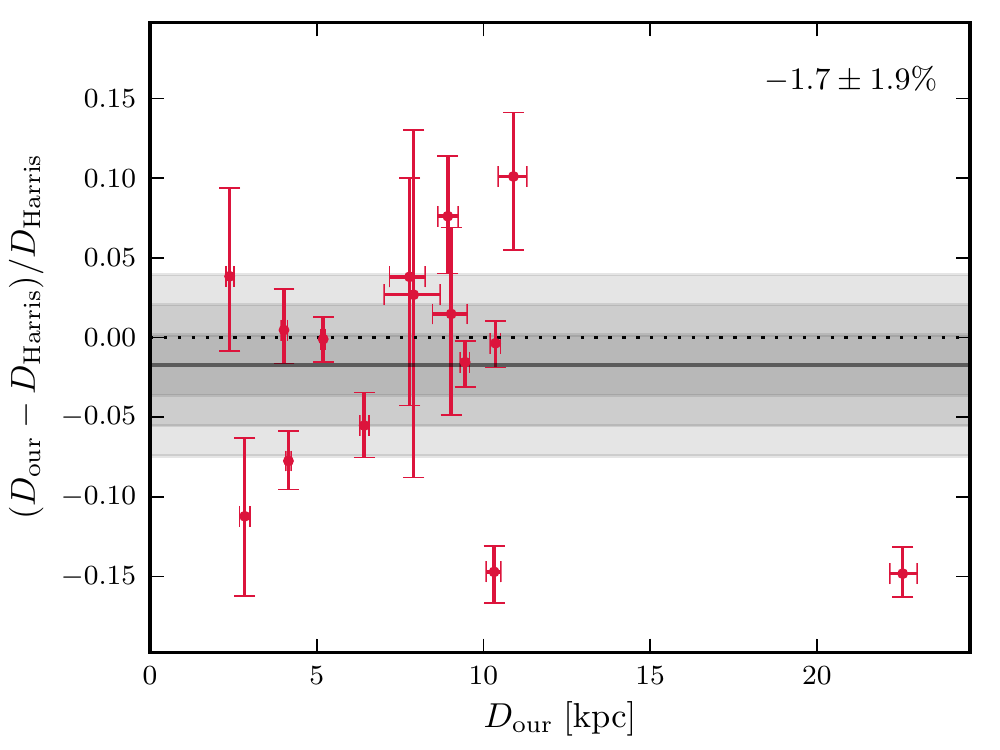}
    \caption{Comparison of our dynamical distance estimates with literature photometric distance estimates from \citetalias{harris1996}. The red points show the fractional difference in distance as a function of our distance estimate for each of our clusters. The dotted line highlights zero, that is perfect agreement between the two estimates. The solid line is the mean of the distance offsets and the shaded areas show the 1-, 2- and 3-$\sigma$ regions, where $\sigma$ is the error on the mean. The mean and error are also shown in the top-right corner of the panel. Overall, the agreement between dynamical and literature distances is very good, suggesting that there are no significant biases in either estimation method.}
    \label{fig:comp_dist}
\end{figure}

In \autoref{fig:comp_dist}, we show the fractional difference between our distance estimates and the \citetalias{harris1996} distance estimates, defined as $\left( D_{\rm our} - D_{\rm Harris} \right) / D_{\rm Harris}$, as a function of our distance estimates. The dotted line marks a difference of zero. The solid line marks the mean of the fractional offsets and the shaded areas mark the 1-, 2- and 3-$\sigma$ regions, where $\sigma$ is calculated as the error on the mean. The mean and $\sigma$ values are shown in the top-right corner of the panel. The most distant cluster in the sample, by some margin, is NGC\,6715 (M\,54); the rest of the clusters lie within 2 and 12~kpc from the sun.

The unweighted mean of the offsets is consistent with zero at $-1.7 \pm 1.9 \%$; this demonstrates that, on average, the dynamical and photometric estimates are in very good agreement over the whole sample. Although some clusters do show considerable disagreement between the dynamical and photometric estimates, this should be further investigated on a cluster-by-cluster basis. We see no evidence of a systematic offset biasing one method to over- or underpredict compared to the other. Furthermore, neither the sign nor the magnitude of the offset correlate with cluster distance.

\subsection{Comparison with literature dynamical distance estimates}
\label{sect:dyn_comp}

For five clusters in our sample, dynamical distances have previously been estimated, so we now compare our estimates to these. As we shall see, overall our estimates are consistent with the earlier studies:

\textit{NGC\,104}. \citet{mclaughlin2006} estimated a dynamical distance of $4.0 \pm 0.35$~kpc, which is in good agreement with our estimate of $4.15 \pm 0.08$~kpc.

\textit{NGC\,5139 (\wcen)}. There have been three dynamical estimates for \wcen: $4.8 \pm 0.3$~kpc \citep{vandeven2006}; $4.73 \pm 0.09$~kpc \citep{vandermarel2010}; and $4.59 \pm 0.08$~kpc \citep{watkins2013}. All three studies attempted to account for the inner complexities of the cluster (rotation, anisotropy, flattening), which may explain why they are more similar to each other than to our estimate of $5.19^{+0.07}_{-0.08}$~kpc (which actually agrees better with the \citetalias{harris1996} photometric distance of $5.2$~kpc).

\textit{NGC\,6266 (M\,62)}. \citet{mcnamara2011} estimated a dynamical distance of $7.054 \pm 0.583$~kpc, which is consistent with our estimate of $6.42 \pm 0.14$~kpc at the 1-$\sigma$ level.

\textit{NGC\,6397}. \citet{heyl2012} estimated a dynamical distance of $2.2^{+0.5}_{-0.7}$~kpc, which is consistent with our estimate of $2.39^{+0.13}_{-0.11}$~kpc.

\textit{NGC\,7078 (M\,15)}. There have been two previous dynamical distance estimates for this cluster: $10.3 \pm 0.4$~kpc \citep{vandenbosch2006}; and $9.98 \pm 0.47$~kpc \citep{mcnamara2004}. Our estimate of $10.36^{+0.15}_{-0.16}$~kpc is in good agreement with both.

\subsection{Comparison with literature mass-to-light ratio estimates}
\label{sect:ml_comp}

We also wish to compare our mass-to-light ratio estimates with those already in the literature. For this comparison, we use the mass-to-light ratios from \citet{mclaughlin2005}, which were estimated from population-synthesis models.  They adopted cluster metallicities from \citetalias{harris1996} and a common age for all clusters of $13 \pm 2$~Gyr. Combining stellar population models for a given metallicity and age with an assumed initial mass function (IMF), they were able to estimate intrinsic colours and average mass-to-light ratios. They provided 6 different estimates using different stellar-population (SP) codes and assuming different IMFs; to begin, we adopt the values calculated using the code presented in \citet{bruzual2003} with a \citet{chabrier2003} disk IMF, as they did for their final analysis. These M/L estimates are included in \autoref{table:results}.

\begin{figure}
    \centering
    \includegraphics[width=\linewidth]{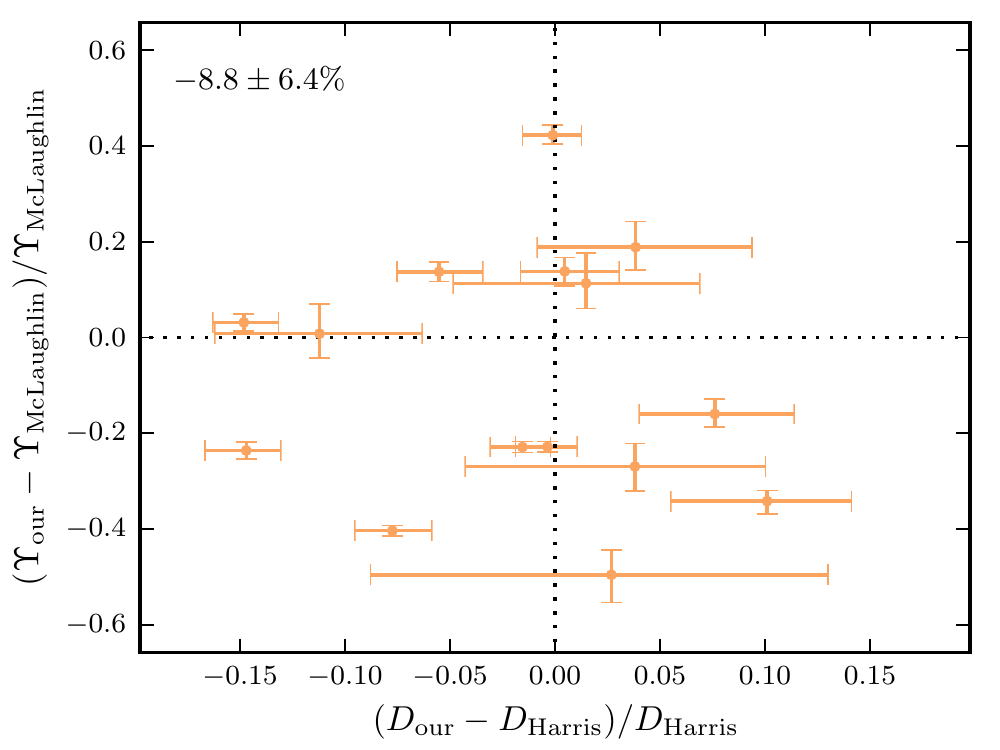}
    \caption{Comparison of dynamical distance and dynamical M/L estimates with literature photometric distances \citepalias{harris1996} and population-synthesis M/L estimates \citep{mclaughlin2005}. The orange points show the fractional offsets of our estimates from the literature values; the dotted lines highlight zero, or perfect agreement. The unweighted mean and error on the mean for the M/L offsets is shown in the top-left corner of the panel. The corresponding value of the distance offsets was already show in \autoref{fig:comp_dist}. The key point here is that the points are scattered across the whole figure, there is no correlation between the direction and magnitude of the distance and M/L offsets.}
    \label{fig:comp_distml}
\end{figure}

As we did for the distances, we calculate the fractional difference between our mass-to-light ratio estimates and the \citet{mclaughlin2005} mass-to-light ratio estimates, defined as $\left( \Upsilon_{\rm our} - \Upsilon_{\rm McLaughlin} \right) / \Upsilon_{\rm McLaughlin}$. In \autoref{fig:comp_distml}, we show the fractional-distance differences and the fractional mass-to-light differences. The dotted lines mark offsets of zero. The unweighted mean M/L offset and error on the mean are shown in the top-left corner of the panel. (The mean distance offset is already shown in \autoref{fig:comp_dist}.)

The unweighted mean of the M/L offsets is consistent with zero at the 1.3-$\sigma$ level: $-8.8 \pm 6.4 \%$. It is clear from the figure and from the means and errors that the fractional mass-to-light differences are larger, in general, than the fractional distance differences. The mass-to-light offsets are typically 20-40\%, while the distance offsets are typically within 10\%. However, the scatter in the plot shows no correlation. The direction and magnitude of the distance offset is independent of the direction and magnitude of the M/L offset, thus indicating that our method is robust. Moreover, the good agreement of our M/L estimates with the \citet{mclaughlin2005} estimates implies that a \citet{chabrier2003} disk IMF is consistent with our data for all clusters.

\subsubsection{Initial mass functions}

As previously mentioned, \citet{mclaughlin2005} provided 6 different M/L estimates using different SP codes and different IMFs. We now compare our dynamical M/L estimates with each of these population-synthesis estimates. Once again, we calculate the fractional offset of our measurements from the population-synthesis estimates and calculate the mean and the standard error on the mean. The models are summarised and M/L offsets are reported in \autoref{table:mlmodels}.

\begin{table}
    \caption{Consistency of different population-synthesis M/L estimates.}
    \label{table:mlmodels}
    \centering
    
    \begin{tabular}{cccc}
        \hline
        \hline
        Model & Code & IMF & $\Delta \Upsilon$ \\
        (1) & (2) & (3) & (4) \\
        \hline \\[-1em]
        1 & BC     & Chabrier (disk) &  -8.8 $\pm$ 6.4 \\[0.4em]
        2 & BC     & Salpeter        & -45.7 $\pm$ 3.9 \\[0.4em]
        3 & PEGASE & Chabrier (disk) & -17.9 $\pm$ 5.8 \\[0.4em]
        4 & PEGASE & Salpeter        & -50.6 $\pm$ 3.6 \\[0.4em]
        5 & PEGASE & Chabrier (GC)   &  -8.6 $\pm$ 6.6 \\[0.4em]
        6 & PEGASE & Kroupa          & -15.0 $\pm$ 6.1 \\[0.4em]
        \hline
    \end{tabular}
    
    \qquad
    
    \textbf{Notes.} Columns: (1) population-synthesis model; (2) SP code used to compute model (BC=\citealt{bruzual2003}, PEGASE=\citealt{fioc1997} v2.0); (3) IMF used for model \citep[sources:][]{chabrier2003, salpeter1955, kroupa1993}; (4) fractional offset of the population-synthesis M/Ls from our dynamical M/Ls. The population-synthesis M/L estimates were all taken from \citet{mclaughlin2005}.
\end{table}

Irrespective of the SP code used to perform the modelling, we see that our dynamical M/L estimates and the population-synthesis estimates using a \citet{salpeter1955} IMF (models 2 and 4) are significantly different. The offsets are at the level of 12-$\sigma$ using the \citet{bruzual2003} code and 14-$\sigma$ using the PEGASE code \citep{fioc1997}. We strongly rule out a \citet{salpeter1955} IMF for Galactic GCs.

The remaining models used a \citet{chabrier2003} disk IMF (models 1 and 3), a \citet{chabrier2003} GC IMF (model 5) or a \citet{kroupa1993} IMF (model 6). All are consistent with our M/L estimates, given the random errors inherent in our analysis and uncertainties due to the modelling and details of the SP codes.

\subsubsection{Mass-to-light ratio variation with metallicity}
\label{sect:ml_feh}

\citet{strader2011} estimated dynamical V-band M/L ratios for a set of 200 globular clusters in M31 and found that their M/L estimates tend to decrease with increasing cluster metallicity. This behaviour directly opposes predictions from population-synthesis models that M/L ratios should increase with increasing cluster metallicity. So let us now consider how our M/L estimates vary with cluster metallicity, which values we take from \citetalias{harris1996}.

\begin{figure}
    \centering
    \includegraphics[width=\linewidth]{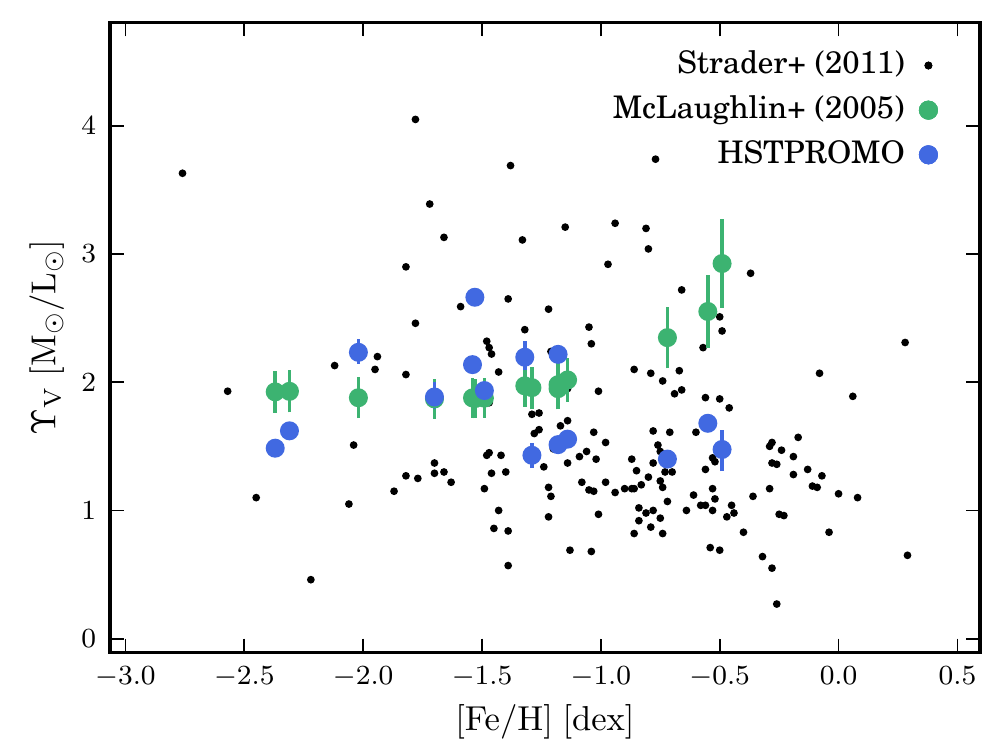}
    \caption{M/L estimates as a function of metallicity. The blue points show our dynamical M/L estimates, with metallicities taken from \citetalias{harris1996}. For comparison, the black points show the dynamical M/L estimates for a sample of M31 GCs from \citet{strader2011}, which tend to decrease with increasing metallicity; the green points show the population-synthesis M/L estimates from \citet{mclaughlin2005} for our clusters, which tend to increase with increasing metallicity. Our estimates are consistent with the \citet{strader2011} M31 clusters over the full range of metallicities but deviate from the population-synthesis estimates at high metallicities.}
    \label{fig:comp_strader}
\end{figure}

In \autoref{fig:comp_strader}, we show our dynamical M/L estimates as a function of cluster metallicity (blue points). For comparison, we show the \citet{strader2011} M31 cluster data (black points) and the \citet{mclaughlin2005} M/L estimates for our clusters (green points, again using metallicities from \citetalias{harris1996}). The distribution of our M/L estimates is consistent with the \citet{strader2011} estimates across the full range of metallicities and exhibits the same decrease in M/L that \citet{strader2011} saw at the metal-rich end of the M31 cluster distribution. Our estimates are also consistent with the \citet{mclaughlin2005} population-synthesis estimates for the metal-poor clusters (as we would expect given the generally good agreement overall discussed in the previous section) but deviate for the most metal-rich clusters.

\citet{strader2011} argued that this behaviour is not caused by dynamical evolution inside the cluster and instead favour a modification to population-synthesis models (in the form of a shallower mass function for metal-rich GCs) to bring the dynamical estimates and population-synthesis estimates into better agreement. However, \citet{shanahan2015} showed that this phenomenon may indeed be dynamical in origin and attribute it to mass segregation in the cluster. It is beyond the scope of this work to investigate further the origin of this phenomenon, however the full PM catalogues are well-suited to studying mass segregation in clusters and this is a topic we plan to address in future papers.

\subsection{Effect of modelling assumptions}
\label{sect:assumptions}

For the modelling part of this analysis, we assumed that our clusters are spherical, isotropic, non-rotating and have a constant mass-to-light ratio. For 12 of our 15 clusters, the models were generally good fits to both the PM and LOS data. This suggests that our assumptions were reasonable for these clusters; we cannot be certain, though it seems convenient that the effect of a number of incorrect assumptions would cancel out so well. However, let us briefly consider what we do know about these clusters and the quality of the assumptions we have made.

\subsubsection{Shape}

The assumption of sphericity implies that quantities of interest vary only with radius and not with position angle so we can consider 1-dimensional radial profiles. We demonstrated in \citetalias{watkins2015} that many of our catalogues contain sufficient data to study 2-dimensional spatial variations in velocity dispersion, however most LOS velocity datasets are too small to obtain 2-dimensional dispersion maps. (Indeed, many LOS velocity datasets are so small that obtaining 1-dimensional profiles can be a challenge.) Furthermore, we require surface brightness profiles for our modelling here, and these are only available as 1-dimensional profiles for most clusters. Deriving 2-dimensional surface-brightness profiles is beyond the scope of this paper. Reassuringly, the 2-dimensional dispersion maps we showed in \citetalias{watkins2015} did not reveal a high degree of flattening in our clusters.

\citet{white1987} provide ellipticities for 14 of the well-fit clusters (only NGC\,288 is missing from their list): the values range from 0.01 for NGC\,6266 to 0.14 for NGC\,5904 and NGC\,6656. In fact, there are four clusters with ellipticities of 0.1 or higher. We might expect the assumption of sphericity to break down for these moderately flattened clusters, however we see no correlation between distance offset and ellipticity. This supports the notion that ignoring ellipticity does not systematically bias the distance estimates.

\subsubsection{Anisotropy}

We showed in \citetalias{watkins2015} that the clusters in our sample are very close to isotropic at their centres. Some clusters do show some radial anisotropy with increasing radius, but the effect is very mild. In most cases, the LOS velocity data extends beyond our PM data, however, even if we extrapolate the anisotropy profiles from \citetalias{watkins2015} to cover the range of the LOS velocity data, the anisotropy is still small. \citetalias{watkins2015} measured projected anisotropy and not intrinsic anisotropy, which is key when considering LOS data alongside our PM data. However, Section 2.7.3 of \citet{vandermarel2010} shows that projected and intrinsic anisotropy are directly linked, so our expectation of isotropy or, at most, mild anisotropy still holds.

Furthermore, we average out the impact of anisotropy by combining both radial and tangential PMs together when calculating our PM dispersion profiles. For steep density profiles, anisotropy will tend to increase one component of motion while decreasing another component of motion by an approximately equal factor so that their mean remains approximately the same. Although this behaviour can break down for flatter density profiles, such as we observe for GCs \citep[eg.][]{cappellari2015}; we performed tests to verify that this is not the case for our clusters. In general, averaging the PMs together does tend to wash out any anisotropy.

Finally, the good agreement between our distances and the Harris photometric estimates suggests that the assumption of isotropy is reasonable.

\subsubsection{Rotation}

We assume that the clusters are not rotating -- this is a necessary assumption because our PM catalogues contain only relative motions and not absolute motions, thus washing out any rotation signatures. As we discussed in detail in \citetalias{watkins2015}, the lack of any rotation signature does not affect the dispersions that we measure. We also know that rotation will generally increase with radius and, as our PM measurements are confined to the central regions, we do not expect significant signal from rotation in our PM data.

For the LOS data sets, rotation has the potential to be more of a concern. The key point to consider here, again, is that we expect little to no rotation at the centre where the LOS data overlaps with our PM data. The good agreement between LOS data and model in the outer regions for most clusters would suggest that rotation is not significant in most cases. However, recall that there were 5 clusters for which we restricted the range of the LOS data to be only the region that overlapped with our PM data due to poor model fits. For these clusters, it is possible that rotation may become significant in the outer regions and should be included for an improved model fit outside of the range of our PM data.

\subsubsection{Mass-to-light ratio}

We assume that mass follows light in clusters, so the mass-density profile is simply a scaled version of the luminosity-density profile; the scale factor is, of course, the M/L ratio. As they evolve, clusters move towards a state of energy equipartition, the effect of which is that massive (bright) stars tend to move more slowly than the low-mass (faint) stars. As a result, the more-massive (brighter) stars sink towards the centre of the cluster and, therefore, have a more concentrated density profile than the less-massive (fainter) stars. This is mass segregation. Owing to the different luminosity-density and mass-density profiles created by this mass segregation \citep{kruijssen2008}, we might expect that the cluster M/L will change with radius.

In principle, our dynamical models could allow the M/L to change as a function of radius, however, for our analysis, we assume that the mass-to-light ratio is constant through the cluster. The M/L at a fixed radius will affect both PM and LOS measurements equally, so a radially-varying M/L is only an issue in cases for which the radial coverage of the PM data and LOS data are different. Once again, the good agreement between model and data, even in cases where the radial coverage of the different data types are not consistent, indicates that the assumption of constant (or nearly constant) M/L through the clusters is reasonable. It is possible that the clusters for which we restricted the range of the LOS data to have an M/L that varies with radius, but the effect is mitigated by considering only the LOS data that overlaps with our PM data.

Further support for the assumption of a constant M/L comes from \citet{vandeven2006} who showed that the mass-to-light ratio of \wcen\ (NGC\,5139) is constant over a wide range of radii. Though it is worth noting that \wcen\ may harbour an intermediate-mass black hole (IMBH) at its centre -- results are conflicting \citep{vandermarel2010, noyola2010}. IMBHs are believed to limit the amount of mass segregation in a cluster because they act as a source of energy at the cluster centre \citep{baumgardt2005, gill2008}; so if \wcen\ does host an IMBH, we would expect little mass segregation and a fairly flat mass-to-light ratio.

IMBHs are not the only mechanism by which mass segregation can be quenched; any source of energy at the centre will achieve the same effect. A significant population of binaries or a segregated population of stellar-mass black holes at the centre of the cluster can also inhibit mass segregation throughout the rest of the cluster \citep[eg.][]{mackey2008}. So there are a number of factors that determine the degree of mass segregation and, hence M/L variation in a cluster.

The degree of segregation within clusters has been studied for individual clusters \citep[eg.][]{pasquato2009, beccari2010} and for large samples of clusters \citep[eg.][]{goldsbury2013}. 14 of our 15 clusters are in the \citet{goldsbury2013} sample; most show some mass segregation but it is relatively mild. A notable exception is NGC\,6388, which shows considerable mass segregation. This may explain why the shape of the Jeans models deviated from the shape of the velocity dispersion profiles, though, recall that the velocity profiles themselves did show good agreement, despite the poor shape of the model.


\section{Conclusions}
\label{sect:conclusions}

We have determined dynamical distance and M/L estimates for 15 Galactic globular clusters. To do this, we combined PM dispersion profiles determined from the \textit{HST} PM catalogs described in \citetalias{bellini2014} with LOS dispersion profiles compiled from literature sources. To these dispersion profiles, we then fitted dynamical models from which we were able to estimate distances and M/Ls. For the majority of our clusters, no PM data was previously available and so this is the first dynamical distance and M/L study performed.

Although the distance estimates themselves are a useful product of this work, the driving force behind this study was to check for consistency between dynamical distance estimates and photometric distance estimates. To this end, in \autoref{sect:dist_comp}, we compared our dynamical distances with photometric distance estimates from \citetalias{harris1996}. A key point in this comparison is the homogeneity of the data analyses that we compared; the data for each cluster is necessarily different as each cluster is unique, however, both here and in \citetalias{harris1996}, the data was treated in a uniform way. This helps to eliminate some possible sources of contention and allows for a more robust comparison.

We showed that the dynamical and photometric distance estimates were consistent with each other within their uncertainties. The mean fractional difference between the two types of estimate was just $-1.9 \pm 1.7 \%$. This indicates that there are no significant biases in either method that may cause one to under- or over-estimate the cluster distances with respect to the other. This also provides an important validation of the stellar evolution theory that underlies photometric distance estimates. Furthermore, neither the sign nor the magnitude of the fractional distance offset correlate with cluster distance.

A side product of this analysis was the estimation of dynamical M/L values for each cluster. We compared these against the population-synthesis M/L values determined in \citet{mclaughlin2005}. Pleasingly, the values agreed within 1.3-$\sigma$ (the mean fractional M/L offset was $-8.8 \pm 6.4 \%$), and we detected no correlation between our M/L offsets from literature values and our distance offsets from literature values. This provides a strong indication that our methods were reliable and our results are robust.

Further, we compared our dynamical M/L estimates with a set of population-synthesis M/L estimates \citep[also from][]{mclaughlin2005} which assumed different IMFs: we ruled out a \citet{salpeter1955} IMF at high significance, but were unable to distinguish between a \citet{chabrier2003} disk IMF, a \citet{chabrier2003} GC IMF and a \citet{kroupa1993} IMF. We also found that our M/L estimates tend to decrease with increasing metallicity; this behaviour directly opposes predicts from SP models, but supports a similar trend observed in a sample of 200 M31 clusters by \citet{strader2011}.

\citetalias{bellini2014} provided catalogues for 22 Galactic globular clusters, however we were only able to use 15 clusters for this analysis. Of the clusters we did not include here, two clusters -- NGC\,6362 and NGC\,7099 -- have LOS literature data, however we found that our PM catalogs (after cleaning) were too small to perform a reliable analysis. For these, it is possible that a more tailored cleaning algorithm would yield more usable stars than the one-size-fits-all cleaning we performed here -- this is beyond the scope of this paper but would be an interesting topic for future work.

For the remaining 5 clusters -- NGC\,362, NGC\,6441, NGC\,6535, NGC\,6624 and NGC\,6681 -- there was no LOS data available in the literature that was sufficient for our purposes here. This illuminates an important shift in dynamical cluster studies. Although LOS measurements and PM measurements are both lacking for clusters, until now, lack of PM data has been biggest drawback. However, we are now in a situation where we have some clusters with LOS data but no PM data, and some clusters with PM data but no LOS data. These five clusters would be prime targets for observing campaigns in the near future in order to obtain larger LOS data sets that could augment the kinematic information provided by our PM catalogues.


\section*{Acknowledgements}

LLW wishes to thank Jeremy Heyl, Rodrigo Ibata, Brian Kimmig, Richard Lane, Carmela Lardo, Anil Seth and Antonio Sollima for kindly providing machine-readable versions of their LOS dispersion profiles. Thank you also to Katharine Larson and Anil Seth for sending their unpublished dispersion profiles. We also wish to thank the referee for a thoughtful and very useful report. Support for this work was provided by grants for \textit{HST} programs AR-12845 (PI: Bellini) and GO-12274 (PI: van~der~Marel), provided by the Space Telescope Science Institute, which is operated by AURA, Inc., under NASA contract NAS 5-26555.

This research made use of Astropy\footnote{\url{http://www.astropy.org}}, a community-developed core Python package for Astronomy \citep{astropy2013}.


\appendix

\section{Restricting literature data}
\label{app:loslimits}

\begin{figure*}
    \centering
    \includegraphics[width=0.33\linewidth]{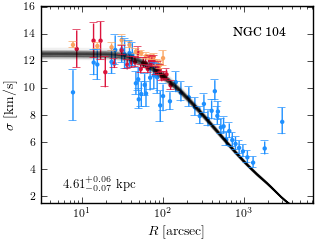}
    \includegraphics[width=0.33\linewidth]{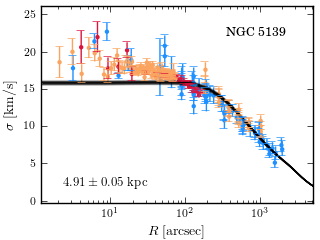}
    \includegraphics[width=0.33\linewidth]{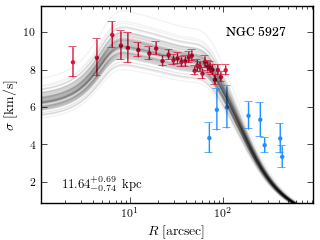}
    \includegraphics[width=0.33\linewidth]{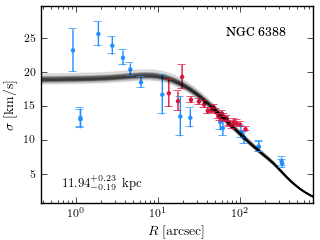}
    \includegraphics[width=0.33\linewidth]{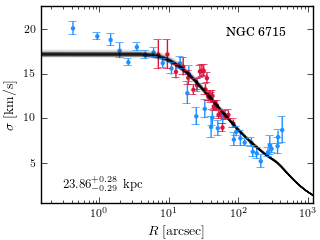}
    \caption{Full fits for restricted clusters.}
    \label{fig:restricted}
\end{figure*}

During preliminary analysis of our clusters, we found that five had several inconsistencies between our PM data, literature LOS data, and the model fit. \autoref{fig:restricted} shows the initial results for these five clusters. As for \autoref{fig:results}, red points show our PM data and blue points show literature LOS data; the black line show the models fits to the red and blue points. Orange points show literature PM data, where available, for comparison only and were not used for fitting. Let us consider each in turn.

\textit{NGC\,104}. Here, the model is a reasonable fit to the LOS data where it overlaps with our PM data, but underestimates the LOS data in the outer regions. This is likely due to the break down of our assumptions in the outer regions of the cluster (see discussion in \autoref{sect:assumptions}). The anomalous two outermost data-points are attributable to evaporation \citep{lane2010}; their inclusion or exclusion does not significantly alter the fit.

\textit{NGC\,5139}. Here, the models do a very poor job at fitting both the LOS dispersions and the PM dispersions. However, as we mentioned in \autoref{sect:models}, \wcen\ is well-known to be anisotropic, rotating and highly-flattened, so it is not surprising that even the best model does not fit the data well. We expect that anisotropy, rotation and flattening will all be more significant and intermediate and large radii, that is, in regions beyond the extent of our PM data.

\textit{NGC\,5927}. The peculiar shape of the surface brightness profile for this cluster (see \autoref{fig:sbprofiles}) gives rise to a peculiar shape for the model dispersion profiles, though the models do seem to trace our PM data fairly well. However, the models fail to accurately reproduce the shape of the LOS dispersion profile outside of the range of our PM data.

\textit{NGC\,6388}. Here, the model is a very poor fit to the data in the central regions of the cluster. This is due to the considerable controversy regarding the shape of the dispersion profile near the centre. Studies of individual radial velocities \citep{lanzoni2013, lardo2015} argue that the dispersion profile stays flat at the centre, while studies using Integral Field Spectroscopy (IFS) measure a dispersion that rises steeply towards the centre \citet{luetzgendorf2011} (as would be expected if the cluster harbours an intermediate-mass black hole at its centre). The PM profile used for the present study does not extend far enough into the centre to comment on either of these claims.\footnote{This will require more careful cleaning of the data and more detailed modelling, and will be the topic of a future paper.}

\textit{NGC\,6715}. For this cluster, the model fails to trace the central rise of the literature LOS dispersion profile or the sharp upturn in the outer regions of the cluster. However, the model is broadly consistent with the LOS data in the region of overlap with our PM data.

In an attempt to improve the fits, we restrict the range of the literature LOS data used for the fitting of these clusters to be only those data points that overlap with our PMs. That is, if our PM dispersion profile has an innermost bin at $R_{\rm in} \pm \Delta R_{\rm in}$ and an outermost bin at $R_{\rm out} \pm \Delta R_{\rm out}$, we restrict the range of the literature data to be $R_{\rm in} - \Delta R_{\rm in} \le R \le R_{\rm out} + \Delta R_{\rm out}$.

For some of these clusters, the inconsistencies are only apparent outside of the range of our PM data, so by restricting the data in this way, we can eliminate problem areas. For others, the fits seem generally poor, but by restricting the range of the LOS data to the region of overlap with our PM data, we hope to at least mitigate any problems outside of the region of our data. As discussed further in \autoref{sect:distances}, the fits are greatly improved for NGC\,104 and NGC\,6715, but still show some inconsistencies for NGC\,5139, NGC\,5927 and NGC\,6388.

It is important to note that restricting the LOS data as we have described here is not the same as fudging the result. In essence, our goal is simply to align the LOS and PM profiles in the radial range where they overlap. Employing dynamical models, as we have done here, allows us to also use the data in the radial range where they do not overlap. This is only improvement if the models are to be trusted and provide a good fit. If they do not fit the data at large or small radii, then this approach actually makes the results worse than when only using the data in the radial overlap region.

\section{Compilation of literature data}
\label{app:lit}

\autoref{table:literature} provides the compilation of literature velocity dispersion profiles used for this paper, as described in \autoref{sect:litdata} and summarised in \autoref{table:litsource}. This table contains profiles for all clusters together. The columns list: the name of the cluster; the radial coordinate (and uncertainty, where available); the velocity dispersion, with uncertainty, of the stars in the bin; the type of data (PM or LOS); and the data source.

\begin{table*}
    \centering
    \caption{Compilation of literature dispersion profiles.}
    \label{table:literature}
    
    \begin{tabular}{lrrrrrcc}
        \hline
        \hline
        Cluster & $R$ & $\Delta R$ & $\sigma$ & $\Delta \sigma^+$ & $\Delta \sigma^-$ & type & source \\
        & \multicolumn{2}{c}{(arcsec)} & \multicolumn{3}{c}{(\kms)} & \multicolumn{2}{c}{} \\
        (1) & (2) & (3) & (4) & (5) & (6) & (7) & (8) \\
        \hline
        NGC\,104  &   13.525 &   4.593 & 11.967 & 0.939 & 0.939 & LOS & \citet{gebhardt1995} \\
        &   24.967 &   2.510 & 12.898 & 1.079 & 1.079 & LOS & \citet{gebhardt1995} \\
        &   32.605 &   2.217 & 12.582 & 1.096 & 1.096 & LOS & \citet{gebhardt1995} \\
        &   40.639 &   2.227 & 12.564 & 0.839 & 0.839 & LOS & \citet{gebhardt1995} \\
        &   48.045 &   2.293 & 10.726 & 0.842 & 0.842 & LOS & \citet{gebhardt1995} \\
        &   57.730 &   4.777 & 10.309 & 0.796 & 0.796 & LOS & \citet{gebhardt1995} \\
        &  140.017 &  47.826 & 10.512 & 0.744 & 0.744 & LOS & \citet{gebhardt1995} \\
        &  457.745 & 157.683 &  8.015 & 0.643 & 0.643 & LOS & \citet{gebhardt1995} \\
        &    7.500 &   0.000 &  9.760 & 1.640 & 2.090 & LOS & \citet{mclaughlin2006} \\
        &   15.000 &   0.000 & 11.820 & 1.050 & 1.130 & LOS & \citet{mclaughlin2006} \\
        \hline
    \end{tabular}
    
    \qquad
    
    \textbf{Notes.} Columns: (1) cluster ID; (2-3) radius and uncertainty; (4-6) dispersion and uncertainty; (7) data type; (8) data source.
    
    (This table is available in its entirety in a machine-readable form.)
\end{table*}


\bibliographystyle{apj}

\bibliography{refs}


\end{document}